\documentstyle[11pt,epsf]{article}
\newcommand{\sptwo}{1.6}

\newcommand{\doublespace}{\edef\baselinestretch{\sptwo}\Large\normalsize}

\begin{document}

\begin{center}
{\large \bf QUANTUM INFORMATION SCIENCE FROM THE PERSPECTIVE OF A DEVICE
AND MATERIALS ENGINEER}
\end{center}

\medskip

\begin{center}
{\it S. Bandyopadhyay \\
Department of Electrical Engineering \\
Virginia Commonwealth University \\
Richmond, VA 23284 \\
E-mail: sbandy@vcu.edu \\
Tel: 1+ (804) 827-6275 \\
Facsimile: 1+ (804) 828-4269}
\end{center}

\begin{abstract}
\noindent This chapter presents a bird's eye view of the rapidly burgeoning 
field of quantum information science. It is intended for a materials scientist,
an electrical engineer in the solid state subdiscipline, or a condensed 
matter physicist. Four areas of quantum information science are discussed:
quantum computing, quantum cryptography, quantum teleportation and quantum 
memory. Emphasis 
is placed on the first sub-area since that is where materials science and 
condensed matter physics is expected to have the strongest impact.
\end{abstract}

\bigskip

\doublespace

\section{Introduction}

Quantum mechanics plays a daily role in our lives. Everything that works -
from the television screen to our brain - draws upon the laws of quantum 
mechanics. Little wonder therefore that quantum mechanics can play a central 
role in fashioning computing machinery. The idea of using quantum mechanics 
to build ``quantum computers'' that outpace ``classical computers'' is not all 
that new. The ideas have been 
around for more than 25 years. It is only recently that this field has 
received widespread attention as a result perhaps of rapid advances 
in device and materials technology, particularly nanotechnology. It is 
not all that difficult anymore to construct solid state devices where 
the quantum mechanical wave nature of electrons is  manifested at a relatively
balmy temperature of 77 K, the temperature of liquid nitrogen. It therefore
behooves the device physics community to seriously ponder the art and 
science of quantum informatics.

What then  is  quantum informatics? Examples are always better than precepts
and therefore it is easiest to grasp the nature of quantum computing 
by looking at some examples. 

In classical digital computing (analog computing is another matter; it
is not much used anyway), one is accustomed to the idea of representing numbers
with binary bits and then manipulating the bits according to a given
set of instructions (the program) to carry out a computation. A binary bit
is the most primitive carrier of information. It can have two values: 0 or 1.
The information content of the bit is its value. 

In the 1930s, there was an ongoing effort to identify a {\it mathematical}
model of the computing process which will be independent of how the computer 
(hardware and software) is implemented. Alan Turing,
Alsonso Church, Kurt G\"odel and Emil Post, independently developed
mathematical models of computation that were independent of any 
assumptions regarding how the computer is built. G\"odel 
identified algorithmic (computation) processes with general 
recursive functions (functions that call themselves). Church identified them
with the so-called the $\lambda$-definable functions \cite{church}.
Turing, on the other hand, provided a very physical picture. He visualized the
computing process as the actions of a so-called universal Turing machine. It
consists of an infinite
tape and a head. The tape has an infinite sequence of bits. The head can 
reset any bit to 0 or 1, reset its own internal memory, and then jump
one bit to the left or right, or simply stay where it is. This machine can
execute any classical 
algorithm and therefore model any classical computer \cite{turing}.

It has now been shown that G\"odel's, Church's and Turing's models are 
all equivalent. Roman Maeder \cite{maeder} has written
a simulator showing 
how the computation of any recursive function, that calls itself, can be 
simulated in a deterministic Turing machine, thereby unifying 
G\"odel's model directly with Turing's.

There are two classes of Turing machines: deterministic and probablistic.
We will discuss them briefly.

\subsection{Deterministic Turing machines}

The deterministic Turing machine consists of an infinitely long tape
partitioned into cells, each containing a 0, or a 1. There is a read/write
head that can move left or right by one cell at a time. The head can exist
is one of a finite set of memory states and contains a set of 
instructions (a program) that specify, depending on the current internal
state and the bit currently being read by the head, how the internal
state of the head
will change, whether the bit being scanned will be changed and in which 
direction
the head will move, if it moves at all. 

The so-called Church-Turing thesis  states that any computation can be modeled 
by the universal 
Turing machine. This thesis is the cornerstone of computer science and its 
implication is profound.
This means that expensive supercomputers with several parallel 
processors and inexpensive personal computers, are, in a sense, equally 
powerful. They are both described by the universal Turing machine.
Given enough memory and time, the personal computers can 
do anything that supercomputers can.

\subsection{Probabilistic Turing machines}

While the head of a deterministic Turing machine, in a given state,
reading a certain bit on the tape, has only one (unique) successor state 
available to it, a probabilistic Turing machine has multiple 
successor states available. Which state is ultimately acquired
depends on a random choice. It is obvious then, that the probabilistic
machine does not produce the correct answer with absolute certainty.
If one requires a correct answer with absolute certainty, then there
is an uncertainty in the length of time the probabilistic Turing
machine must run. In fact, there is a trade-off between the time 
it takes to produce an answer and the probability that the answer
is correct. Curiously, many problems can be solved more efficiently
on a probablistic Turing machine than on a deterministic Turing 
machine using the so-called random algorithms. It has been shown
that anything computable by a probabilistic Turing machine is
computable by a deterministic Turing machine and vice versa,
although the probabilistic machine may be more efficient in 
some cases \cite{gill}. 

\section{Quantum Turing machines}

Despite the immense success of the universal Turing machine in
modeling computation, it has one serious shortcoming. It is 
really not completely universal since it is now understood
that it is not completely independent of the physical model 
of the computer. Perhaps unknowingly, Turing based his machine 
on the models of the classical world. The Turing machine is 
unable to capture the subtleties of the quantum mechanical world
and computation processes based on quantum mechanics. There 
is now a model of {\it Quantum Turing machines}. It has a long
history and started with Charles Bennet first showing that 
there exists a {\it reversible Turing machine} such that 
if the state of the machine is known at any time, it is 
possible to predict the states at all future times and all past times.
There is no dissipation and loss of information in this machine \cite{bennet}.

Bennet's reversible Turing machine was completely classical, but it
portended a quantum mechanical sister since the dynamical evolution of 
an isolated quantum system is also completely reversible in time. In 1980,
Paul Benioff devised a quantum system whose actions mimicked that of
a classical reversible Turing machine \cite{benioff}. Benioff's machine was not 
a true
quantum Turing machine however, since although in between computation 
steps the machine existed in a quantum mechanical state, it reverted
to a classical state at the conclusion of each computation
step. Thus, this machine was no more powerful that Bennel's classical
reversible Turing machine.

In 1982, Richard Feynman showed that no classical Turing machine
could simulate quantum phenomena without an exponential penalty 
in processing time \cite{feynman}. Finally, in 1985, David Deutsch
devised the first true quantum Turing machine. Its read, write and move
operations were executed via quantum mechanical interactions. More importantly,
whereas a classical Turing machine (including the reversible one) could
encode only a 0 or a 1 in each cell of the tape, the quantum machine could 
encode a {\it superposition} of 0 and 1 simultaneously in each cell.
Thus, the quantum machine could encode many inputs simultaneously on
the same tape and perform a calculation on all of them in one step,
completing the calculation  in the time it takes 
for a classical machine to do the calculation on only one of the inputs.
This immense speed up is called {\it quantum parallelism}. More on it later.  

\section{Qubits}

To understand quantum parallelism, and indeed to understand the nuances
of quantum computing, one must first understand the concept of quantum 
bits (or ``qubits'').

Quantum computing does not process classical bits. Instead, it processes
quantum bits (or ``qubits'') which are neither 0 or 1, but a coherent 
superposition of both 0 and 1. This means that the bit is neither 0, nor 1,
which does not make sense classically, but makes perfect sense quantum
mechanically. The qubit can be written as 
\begin{equation}
qubit = a_0|0> + a_1|1>
\end{equation}
where $|0>$ denotes the state in which the qubit has a value of 0 and 
$|1>$ denotes the state in which the qubit has a value of 1. The coefficients
$a_0$ and $a_1$ are complex quantities whose squared magnitudes denote the 
probability that if a measurement is performed on the qubit, it will be 
found to have a value of 0 and 1 respectively. Note that although a qubit
can exist in suspended animation -- between a 0 and a 1 -- 
it must ultimately succumb to the fate of having a definite classical value (0 
or 1) 
when it is ``measured''. The outcome of the measurement is unambiguous.
However, prior to the measurement, the qubit is neither 0, nor 1. It is
in a so-called coherent superposition state, and the measurement can yield
a value of either 0 or 1. All  we can say is that the probabilities of 
getting those two values are respectively $|a_0|^2$ and $|a_1|^2$. Note also
that since the measurement can yield a value of only 0 or 1, and nothing else,
it follows that
\begin{equation}
|a_0|^2 + |a_1|^2 = 1
\end{equation}
We emphasize that Equation (1) {\it does not} imply that the qubit is 
sometimes in state $|0>$ with probability $|a_0|^2$, and the rest of the time in
state $|1>$ with probability $|a_1|^2$. It is only after measurement that the
qubit collapses to a classical bit and assumes a definite value of 0 or 1. Prior 
to
the measurement, it {\it does not} have a definite value; it is both 0 and 1, 
{\it all the time}. It is therefore called a superposition state.

While one can easily understand the meaning of ``superposition'' from the 
foregoing discussion, to understand 
the implication of a ``coherent'' superposition requires a little more 
reflection. Coherence has to do with the phases of $a_0$ and $a_1$ (being
complex quantities, they have an amplitude and a phase). It might appear
that the phase is never relevant since these coefficients seemingly determine 
only the probabilities which, in turn, depend only on the squared magnitudes. 
This is actually not true. Consider a hypothetical bit-flip operator 
$\hat{0}_{flip}$
whose action is to flip a bit from 0 to 1 and vice versa (a classical 
realization of a bit flip operator is an inverter or NOT gate). Its action 
on a bit can be represented as 
\begin{eqnarray}
\hat{0}_{flip}|0> = 1|1> \nonumber \\
\hat{0}_{flip}|1> = 1|0>
\end{eqnarray}
Note from the above that $0>$ and $|1>$ are not eigenstates of the operator.
If we want to calculate the expected value of this operator for a system
described by a qubit, we will get (following the usual prescription of 
quantum mechanics)
\begin{eqnarray}
expected~value & = &  <qubit|\hat{0}_{flip}|qubit> \nonumber \\
& = & |a_0|^2<0|\hat{0}_{flip}|0>
+ |a_1|^2<1|\hat{0}_{flip}|1> + a_0*a_1<0|\hat{0}_{flip}|1> + 
a_1*a_0<1|\hat{0}_{flip}|0> \nonumber \\
& = & a_0*a_1 + a_1*a_0 \nonumber \\
& = & 2|a_0||a_1|cos \theta
\end{eqnarray}
where $\theta$ is the phase angle difference between $a_0$ and $a_1$. In 
deriving the last equation, we used Equations (1) and (3) and also the fact 
that $|0>$ and $|1>$ are orthonormal states by themselves. The fact that 
$|0>$ and $|1>$ are orthogonal to each other is obvious because any
measurement outcome can produce either a 0 or a 1, and absolutely nothing in 
between.
Thus the two possible outcomes are mutually exclusive, meaning that the 
two states are orthogonal to each other.

Equation (4) clearly shows that there are quantities that depend on the 
phases of $a_0$ and $a_1$. Thus maintaining the correct phase relations
between these quantities, or ``coherence'' is important. Quantum computation
depends critically on this coherence.

Before ending this section, we point out that $a_0$ and $a_1$ are 
continuous (analog) variables whose magnitudes can take any value 
between 0 and 1 and whose phases can also take any value between 0 and 2$\pi$.
On the other hand, the states $|0>$ and $|1>$ correspond to digital binary bits. 
Thus, quantum computation is neither quite analog computing, nor digital,
but something in between.

\section{Superposition states}

We are now ready to provide the first glimpse of the power of 
quantum computing. Consider a bistable atom. We can use the excited state 
to encode the binary bit 1 and the ground state to encode the binary bit 0.
Classically, this atom can store only one bit of information (whose value
can be 0 or 1) because we have two degrees of freedom (ground and excited)
and two possible values of the binary bit. Since each atom can store only one 
bit,
$N$ bistable atoms can store $N$ binary bits classically. But now,
if we can create a coherent superposition of the states of  two atoms,
the corresponding qubit will be written as
\begin{equation}
qubit_{2~atom} = a_{00}|00> + a_{01}|01> + a_{10}|10> + a_{11}|11>
\label{qubit}
\end{equation}
where the state $|ij>$ corresponds to the first atom being in state $|i>$ 
and the second atom in state $|j>$. This system can, in principle, store
2$^2$ bits of information corresponding to (i) both atoms being in the 
ground state, (ii) first in ground state and second in excited, (iii) first in 
excited and second in ground, and (iv) both excited. Note that the classical
case would correspond to only one of the coefficients in Equation (\ref{qubit})
being non-zero. Extending this to $N$ bistable atoms, we can write the 
corresponding
qubit as
\begin{equation}
qubit_{N-atom} = \sum_{x_1x_2...x_N}a_{x_1x_2...x_N}|x_1x_2...x_N>
\end{equation}
where the $x$'s can take the value 0 or 1. The above equation has 2$^N$ terms.
It means that while $N$ bistable atoms can store $N$ binary bits classically,
the same system can store 2$^N$ classical bits quantum mechanically. Here is the
clincher. If we want to store 2$^{1000}$ binary bits of information, it is
impossible in the classical realm since the number 2$^{1000}$ is larger than the 
number of atoms in the known universe. However, in the quantum mechanical 
realm, just 1000 bistable atoms in a coherent superposition state can do the 
job. These 1000 atoms could store more information than all the hards disks
in the universe made over the life of the universe!

Now the tempered vision. One might be tempted to think that we can make
a huge ``memory'' storing 2$^N$ binary bits of information using 
just $N$ bistable atoms. This point of view would be {\it incorrect}.
A true memory should be such that every time it is accessed, it returns
the {\it same} data which is stored in it. For instance, if we have stored 
the binary string 11001, we should get this string {\it every time}  the memory
is read. But the quantum memory we 
are talking about does not satisfy this requirement. Every time we read the 
memory, the qubits collapse to classical bits and we only get $N$ 
(not 2$^N$) bits out of it. More importantly, the values of these bits 
change with each measurement because of the probabilistic nature of 
quantum measurement. It is never the same bit string in two successive
measurements! Surely, this cannot be judged a reliable memory. There is 
a subtle difference between {\it memory} and {\it storage}. Memory 
implies accessibility and fidelity, storage does not. We may store 2$^N$ bits of 
information, but not be able to access them. However,
the ability of the quantum system to co-exist in 2$^N$ states is what is 
important. It forms the basis of an attribute known as quantum parallelism.

\section{Quantum Parallelism}

In the world of classical computers, parallelism refers to parallel 
(simultaneous) processing of different information in different processors. 
Quantum 
parallelism refers to simultaneous processing of different information 
or inputs in the {\it same} processor. This idea, due to Deutsch, refers
to the notion of evaluating a function once on a superposition of 
all possible inputs to the function to produce a superposition of outputs.
Thus, all outputs are produced in the time taken to calculate one 
output classically.
Of course, not all of these outputs are accessible since a measurement
on the superposition state of the output will produce only 
one output. However, it is possible to obtain certain joint properties
\cite{josza} of the outputs, and that is a remarkable possibility.

Let us exemplify quantum parallelism more concretely.
Consider the situation when $N$ inputs $x_1$, $x_2$, ... $x_N$ are provided to a 
computer and their functions $f(x_1)$, $f(x_2)$, ... $f(x_N)$ are to be 
computed.
The results are then fed to another computer to calculate the functional
$F(f(x_1), f(x_2), ... f(x_N))$.

With a classical computer, we will calculate $f(x_1)$, $f(x_2)$, ... $f(x_N)$ 
serially,
one after the other. With a quantum computer, the story is different.

Prepare the initial state as a superposition of the inputs.
\begin{equation}
|I> = {{1}\over{\sqrt{N}}}(|x_1> + |x_2> + ... + |x_N>)
\end{equation}
Let it evolve in time to produce the output
\begin{equation}
|O> = {{1}\over{\sqrt{N}}}(|f(x_1)> + |f(x_2)> + ... + |f(x_N)>)
\end{equation}
Note that $|O>$ has been obtained in the time required to perform a
single computation. Now, if $C$ =  $F(f(x_1), f(x_2), ... f(x_N))$,
can be computed from $|O>$, then a quantum computer will be of great
advantage. This is an example where ``quantum parallelism'' can be
used to speed up the computation tremendously.

There are two questions now. Can $C$ be computed from a knowledge of
the superposition of various $f(x_i)$ and not the individual $f(x_i)$s?
The answer is ``yes'', but for a small class of problems. These are
called the Deutsch-Josza class of problems which can benefit from quantum 
parallelism.
Second, can $C$ be computed correctly with unit probability. The answer is 
``no''. $C$ cannot be computed with unit
probability. However, if the first answer is wrong (hopefully, the computing 
entity can differentiate right from wrong answers), then the experiment
or computation is repeated until the right answer is obtained. The probability
of getting the right answer within $k$ iterations is $(1 - p^k)$ where 
$p$ is the probability of getting the wrong answer in any iteration.
The mean number of times the experiment should be repeated is
$N^2 - 2N -2$.

The following is an example of Deutsch-Josza class of problems.
For integer $0 \leq x \leq 2L$, given that the function $f_k(x)$ $\in$ [0,1] has 
one of two properties -- (i) either $f_k(x)$ is independent of 
$x$, or (ii) one half of the numbers $f_k(0), f_k(1),....f_k(2L-1)$ are
zero -- determine which type the function belongs to using the fewest
computational steps. 

The most efficient classical computer will require $L+1$ evaluations,
whereas according to Deutsch and Josza, a quantum computer can solve this 
problem with just two iterations.

\section{Classical and Quantum Complexity}

In computer science, complexity refers to how efficiently a problem can 
be solved.  A measure of the efficiency is
the rate of growth of the time or memory required to solve a problem 
as the size of the problem increases. Complexity measures are independent
of the make and prowess of the computer, but depend on the mathematical 
model of the computer, such as whether it is a deterministic classical
Turing machine, a probabilistic classical Turing machine, or a quantum 
machine. In fact, many problems that are deemed intractable on a deterministic 
classical Turing machine can be solved very efficiently with high probability of 
success on a probabilistic classical Turing machine. In classical 
complexity theory, there are 5 classes of problems \cite{collins}. They are 
shown in Table 1.

The study of quantum complexity classes begun with Deutsch \cite{deutsch1}.
Since the quantum Turing machine is a quantum mechanical refinement of 
probabilistic Turing machines, the quantum complexity classes are close
cousins of the probabilistic complexity classes. There are three known
quantum complexity classes \cite{collins} which are shown in Table 2.

\bigskip

\begin{table}
\caption{Classical complexity classes}
\bigskip
\begin{tabular}{|l|l|l|}
\hline
{\bf Class} & {\bf Description} & Example \\
\hline
P & The running time of the & Division, addition \\
& algorithm is a polynomial & \\
& in the size of the input & \\
&& \\
NP & Non-deterministic polynomial & Factoring integers \\
& running time & \\
&& \\
NP-complete & A subset of problems & Traveling salesman \\
& in NP that can be mapped & problem \\
& into one another in & \\
& polynomial time. If one & \\
& of the problems is tractable, & \\
& they all are & \\
&& \\
ZPP & Can be solved with certainty & \\
& by probabilistic Turing & \\
& machines in polynomial & \\
& time on the average & \\
&& \\
BPP & Can be solved in & \\
& polynomial time with & \\
& probability of correctness $>$ 2/3 & \\
&& \\
\hline
\end{tabular}
\end{table}

\bigskip

\begin{table}
\caption{Quantum  complexity classes}
\bigskip
\begin{tabular}{|l|l|l|}
\hline
{\bf Quantum Class} & {\bf Description} & Comparison with Classical \\
\hline
QP & The running time of the & P in included in QP. \\
& algorithm is a polynomial & Quantum computer can \\
& in the size of the input & solve more problems \\
& and the problem can be  & in polynomial time \\
& solved with certainty & than its classical counterpart\\
&& \\
BQP & Can be solved in  & For this class, quantum \\
& polynomial running time & machines are at least \\
& with probability of success $>$ 2/3 & as powerful as classical\\
&& ones. It is not known \\
&& if they are more powerful \\
&& \\
ZQP & Can be solved with & ZPP is included in ZQP. \\
& zero error probability & Quantum computers can \\
& in polynomial time & solve more problems in this\\
&& class than classical computers \\
&& \\
\hline
\end{tabular}
\end{table}

\subsection{History of quantum complexity theory}

Deutsch's idea of quantum parallelism \cite{deutsch1} was an 
interesting concept, but certainly did not establish the 
superiority of quantum computers over classical computers.
With quantum computers,
even though one could, in principle, calculate the outputs
corresponding to all possible inputs in one go, when the
final answer is read, only one of the outputs is obtained
since the superposition state (wave function) collapses to 
a classical state under the read operation which constitutes
a measurement. Worse still, all the other outputs are 
permanently lost.

In 1992, Deutsch and Josza \cite{deutsch2} contrived a problem
that a deterministic Turing machine could solve in linear 
time, but a quantum Turing machine solved in a time that 
was a polynomial in the logarithm of the problem size. Thus,
the quantum machine was exponentially more efficient than the
deterministic machine. However, the probabilistic machine
was equally efficient as the quantum machine and hence the
superiority of the quantum over the classical was still 
elusive. 

The following year, Bernstein and Vazirani came up with 
the first problem where the quantum Turing machine beat 
{\it both} the probabilistic and the deterministic classical
Turing machines. The problem was to sample from the Fourier
spectrum of a Boolean function of $n$ bits. 

The crowning achievement in this area was the work of 
Peter Shor \cite{shor1} who discovered a polynomial 
time algorithm for factoring large integers. This 
is the first problem of great practical significance 
that a quantum computer can solve in polynomial time
(in classical complexity class, this is an NP problem).
The practical significance is associated with breaking
secure cryptographic codes. Many cryptosystems are
based on {\it trapdoor functions}. These are functions
that are easy to compute, but their inverse is difficult 
to compute. The best example is the following.
It is relatively easy to find the product of two
large prime integers (multiplication is in the complexity
class P), but the reverse process of 
factorizing the product back into two prime
integers is unbelievably  difficult (it is in the class
NP). Williams and Clearwater, in their very popular
book on quantum computing, give the following example \cite{collins}.
Consider two prime numbers 15485863 and 15485867. Multiplying
these numbers by hand using pen and paper took this
author a few minutes to produce the product
239812014798221. The author would not have been 
able to factorize the product back into the prime 
factors in his lifetime (of course, he would have run out 
of patience long before his life ended).

What is true of humans is also true of computers.
The most popular classical algorithm for factoring 
is the Multiple Polynomial Quadratic Sieve for numbers 
with 100-150 decimal digits \cite{silverman}. The running 
time of this algorithm grows sub-exponentially, but super-polynomially
in time. In 1978, Rivest, Shamir and Adelman challenged 
computer scientists to factorize an integer
consisting of 129  digits. The challenge 
was finally met in 1994 by a team of computer scientists
using over 1500 networked computers. Crandall has 
estimated that this task would require a computer
executing 1 millions instructions per second, 5000 years
to compute \cite{crandall}.

Trapdoor functions are ideal for use in cryptosystems.
Suppose I want to transmit a secret message to my friend
and I encode the message in the prime number 15485863 using
a publicly known encryption method. I have also 
provided my friend with the prime number 15485867 as a ``key''
ahead of time.
I then transmit the {\it product} of the message and the key - in this case, the 
number 239812014798221 - over  a public
channel to my friend. For me, multiplying the two numbers
and checking their primality takes very little time.
Upon receipt, my friend divides his key into the number
239812014798221 and extracts the actual message 15485863 quite easily. An 
eavesdropper who does not have the key gets only the 
number 239812014798221 and has to factorize it into its prime constituents. 
That is hard
and takes much longer than the period over which the message 
has any usefulness to the eavesdropper.

This, of course, is a very primitive cryptosystem.
There are more sophisticated 
cryptosystems such as the RSA scheme devised by
Rivest, Shamir and Adelman \cite{rivest}. It
is similar in spirit to what was just described
and relies on the difficulty of factorization 
for security. Hence, the development of an 
algorithm that can factorize large numbers
in polynomial times threatens the security of 
many of our most advanced cryptosystems. Hence,
Shor's algorithm is a very important advance.

Shor's algorithm relies on a result from number theory
that establishes a relationship between the period of 
a particular periodic function to the factors of an 
integer. Let us say that we wish to factorize an 
integer $n$. We will construct a function $\phi_n(x)$ = 
$r^x mod (n)$ where $r$ is an integer chosen at random
that is co-prime to $n$, meaning that the greatest
common divisor of $r$ and $n$ is 1. As we increase the 
argument $x$ of the function $\phi_n(x)$ in steps,
taking $x$ = 0, 1, 2, .. etc., the function $\phi_n(x)$
ultimately fall into a periodic pattern. The period 
can be used to find the factors of $n$ efficiently, but there
is unfortunately no efficient classical algorithm to 
find the period in the first place. Shor's contribution
was to develop a quantum algorithm for finding this period
efficiently. His technique relies on quantum parallelism
to find the period of the function $\phi_n(x)$ rather quickly.
Another factorizing algorithm for quantum computers has been proposed by Kitaev
\cite{kitaev}.

Similar ideas have been used to solve another problem,
namely database search. Grover \cite{grover} has discovered
an algorithm for finding a single item in an unsorted 
database in square root of the time it will take on a 
classical computer. These ideas have been extended 
to find the  minimum \cite{durr}
of the database as well, more efficiently than any classical 
computer can.

Unfortunately, at this time, there are only {\it two}
classes of algorithms - factorizing and searching databases - 
where quantum computers evince an edge. No significant 
advancement has been made in finding other classes of algorithms over the last
five years, to the best of the author's knowledge.
Since the field is so new, there is hope however that rapid 
advances may be around the corner.

\section{The early history  of quantum information science: reversible 
computers}

Quantum computing has interesting roots. Its progenitor, reversible computing,
itself has a long history and is extremely relevant to today's nanoelectronics.
There is a long standing belief in the device research community that the 
silicon steamroller, that paved the way for the microelectronics revolution,
will ultimately run out of steam. Future silicon CMOS may encounter 
insurmountable problems due to excessive power dissipation, breakdown of scaling 
laws and certain fundamental limits 
imposed by the laws of quantum mechanics. While  the fundamental limits are
still a moot issue, excessive power dissipation is universally acknowledged to 
be a serious problem.
Microelectronic logic gates of present-day vintage dissipate about 0.1 pJ
of energy per switching cycle.  The Semiconductor Industry Association's
National Technology Roadmap projects that by the year 2007, the 
dynamic power dissipated in CMOS devices  will be 
600 nW per logic gate with a gate density of 5$\times$10$^{7}$/cm$^{2}$,
corresponding to a dissipation of 30 W/cm$^{2}$ of chip area \cite{sia}. 
$\footnote{A good toaster oven dissipates only about 100 W/cm$^2$. None of this
is of course removed. It is used to toast bread and bagels.}$
Fifteen years
ago, removal of 1000 W/cm$^{2}$ was demonstrated in a silicon chip 
\cite{tuckerman}.
Unfortunately, heat sinking technology is not keeping pace with  
solid state circuits technology and excess heat removal will soon be 
a problem. Consequently, it appears that at least in the near future, the gate 
density may be constrained to 10$^{10}$ 
gates/cm$^{2}$ from mere heat sinking considerations. Any denser device density 
will require either more efficient
heat removal techniques, or less power dissipation per logic
gate. It is the latter objective that has captured the imagination of 
device physicists.

In a seminal paper published in 1961 \cite{landauer}, Rolf Landauer addressed 
the fundamental issue of dissipation and showed that the minimum 
energy that must be dissipated in a single {\it logically irreversible} bit 
operation is $kTln2$ ($k$ = Boltzmann constant and $T$ = absolute temperature
of the logic device)
which is about 4$\times$10$^{-21}$ Joules at room temperature (see Section 7.1 
for a derivation). We will explain ``logical irreversibility'' later, but first 
note that this figure is far smaller than what CMOS or single electron 
transistors \cite{korotkov_ssdm} will dissipate in a logic bit operation by the 
year 2007. However, the very existence of this figure portends a practical limit 
to downscaling of conventional logic circuits. Assuming that the most advanced 
devices, constrained only by the 
Landauer limit, will switch in 1 picosecond, the power dissipated will be
4 nW/gate. Moreover, assuming that heat sinking technology will allow removal
of only 10 kW/cm$^{2}$, the gate density will saturate to 2.5$\times$10$^{13}$
gates/cm$^{2}$ unless dramatic improvements in heat sinking are achieved.
The alternative is to seek ways to circumvent the $kTln2$ barrier.

Before we address ways of overcoming the $kTln2$ barrier, we need to 
derive the existence of this barrier first because it will also show
us how we can circumvent it.

\subsection{Derivation of the ``kTln2'' limit}

From simple thermodynamics, the energy dissipated when a system is 
switched from one state to another is
\begin{equation}
\Delta E = k(T\Delta S + S \Delta T)
\label{landauer}
\end {equation}
where $k$ is the Boltzmann constant, $T$ is the absolute temperature 
and $S$ is the system's entropy.

The entropy is given by
\begin{equation}
S = \sum_i p_i ln(p_i)
\end{equation}
where $p_i$ is the probability of the system to be in the $i$-th state.
For a binary device, there are only two states ($i$ = 0,1).

Let us consider a device which is switched. Prior to switching, it had 
equal probability of being in the state 0 and 1. Therefore, its 
initial entropy is 
\begin{equation}
S_{initial} = p_0ln(p_0) + p_1ln(p_1) = 0.5ln(0.5) + 0.5ln(0.5) = -ln2
\label{entropy}
\end{equation}
After switching, we can measure the final state and know for sure whether
the system is in state 0 or 1. Let us say that we find  it is in state 1. 
Then,
\begin{equation}
S_{final} = 0ln(0) + 1ln(1) = 0 
\end{equation}
Therefore, the change in entropy is 
\begin{equation}
\Delta S = S_{final} - S_{initial} = ln2
\end{equation}

For an isolated system $\Delta T$ $\geq$ 0. Substituting all this in
equation \ref{landauer} yields
\begin{equation}
\Delta E_{min} = kTln2
\end{equation}

\subsection{Physical and logical irreversibility}

This derivation tells us a lot. Initially, the system could have been in
either state 0 or 1. Later we found out that it was in the 1 state.
This constricts the phase space for the system's state (decreases the
number of possibilities) and causes 
dissipation. The entropy is raised when we constrict the phase space 
in this fashion. But what if we could always deduce unambiguously
what the initial state was from a measurement of the final state?
In that case, the initial state also has only one possibility, not two.
Consequently, we would not have constricted the phase space and would
have caused no dissipation. It is easy to see from our derivation
that in this case $S_{initial}$ = $S_{final}$ = 0. Therefore, $\Delta S$ = 0 and 
 $\Delta E_{min}$ = 0. To avoid dissipation, the trick is to ensure
that if we {\it reverse} the system in time, it ends up in a {\it unique}
state. Such a system is said to be time reversible, and therefore 
dissipationless. Dissipation 
accrues from time irreversibility, and time irreversibility is also called 
physical irreversibility.

\begin{figure}
\epsfxsize4.2in
\centerline{\epsffile{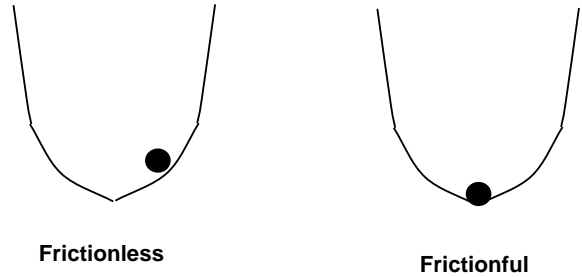}}
\caption{\tt Two wells, one frictionless and the other is not. A marble rolls 
along the walls.
Knowing the 
position and 
momentum of a marble in the 
frictionless well at any instant of time allows us to compute the position and 
momentum at all future and past time. However, for the well with
friction, the final position of the marble is always at the bottom and it is at 
rest. 
We cannot predict from the final position and momentum what the initial position 
and momentum were. \label{wells} }
\end{figure}

Landauer's seminal contribution was to connect physical irreversibility 
to {\it logical irreversibility} and vice versa. This connection can
be visualized better with a concrete example pointed out to the author by a 
colleague \cite{vwani_private}. Consider 
two parabolic wells (Fig. \ref{wells}). One has walls that are frictionless and 
the other 
has normal walls with friction. A marble rolls along both walls. If we 
know the position and velocity of the marble in the frictionless well at any 
given time,
we can always determine the position and velocity at any {\it previous} 
instance of time using Newton's laws which are reversible in time.
But in the other well, the final position of the marble (after a long enough 
time has elapsed) is always at
the bottom and the final velocity is always zero, {\it regardless
of the initial position and velocity}. The second well has dissipation
due to friction and hence we cannot deduce the initial state from the 
final state. If the position and velocity are considered as logic 
variables, the well will be a logic gate, the initial and final states will be 
inputs and outputs. Then, the second well is logically irreversible and causes 
dissipation. The first well is logically reversible and causes 
no dissipation. 

\subsection{Reversible and irreversible gates}

There is a great deal of interest in classical, but reversible, gates
since these gates can overcome the $kTln2$ barrier and be dissipationless.
Irreversible gates, on the other hand, must dissipate at least $kTln2$
amount of energy per bit operation.

Let us look at some examples of (classical) logically reversible  and  logically 
irreversible gates. Consider the simplest  AND gate with two inputs and one 
output whose 
truth table is given below

\bigskip
\begin{tabular}{|l|l|l|}
\hline
Input 1 & Input 2 & Output \\
\hline
0 & 0 & 0 \\
0 & 1 & 0 \\
1 & 0 & 0 \\
1 & 1 & 1 \\
\hline
\end{tabular}

\bigskip

If the output is 1, we can confidently state that the inputs were both 1.
However, if the output is 0, then we cannot state what the inputs were
since there are three sets of possibilities. Because we cannot deduce
the inputs unambiguously from the output alone - there is no unique relation 
between the two - this is a logically irreversible gate. 
On the other hand, it is obvious that an inverter is a logically
reversible gate. 
There is a unique relation between the input and the output; one is always the 
logic complement 
of the other. Therefore, if we know one, we always know the other.

There is another (mathematical) way to view this. An inverter converts
a 0 to a 1 and vice versa. Thus, the 2$\times$2 matrix operator  relating an 
input vector to the output vector
(the vectors span the Hilbert spaces of the inputs and outputs) is
given by the relation:
\begin{eqnarray}
\left[ \begin{array}{c}
             0 \\
             1 \\
\end{array}   \right] 
& = \left[ \begin{array}{cc}
             0 & 1 \\
             1 & 0 \\
\end{array}   \right] &   
\left[  \begin{array}{c}
              1 \\ 
              0 \\
\end{array} \right ] \nonumber \\ 
output & A-matrix & input 
\label{A-matrix}
\end{eqnarray}
The matrix operator for a logic gate is called
an A-matrix. For an inverter, or for that matter any logically reversible
gate, the A-matrix must be invertible and its determinant must not vanish.
It also stands to reason that a reversible gate must have as
many outputs as inputs.

\subsection{Universal reversible gate}

Anybody somewhat familiar with classical Boolean logic circuits knows that  
there are two primitive gates that are 
``universal'' in the sense that any arbitrary logic circuit can be built by
using just (either one of) those gates alone. They are the NAND gate and the NOR 
gate. A fundamental 
theorem of Boolean algebra is that any Boolean function can be 
completely re-written in terms of NAND or NOR functions. Therefore,
any Boolean combinational circuit can be built with just NAND or NOR gates.

Neither  the NAND, nor the NOR gate is however reversible. So the question is: 
are there 
universal reversible gates and if so, what are they? One universal reversible 
gate is the Toffoli-Fredkin gate \cite{fredkin}. It has 3 inputs ($a,b,c$) and 3 
outputs ($a',b',c'$) (remember that the number of inputs and outputs have to be 
equal for a reversible gate). The Toffoli-Fredkin (T-F) gate
is a universal gate in the sense that any reversible computation can be
achieved by repeated operations of the T-F gate. In other words, any
reversible combinational circuit can be built by connecting T-F gates.
The truth table of a T-F gate is
\bigskip

\begin{tabular}{|c|c|c|c|c|c|}
\hline
a & b & c & a' & b' & c' \\
\hline
& & & & & \\
0 & 0 & 0 & 0 & 0 & 0 \\
0 & 0 & 1 & 0 & 0 & 1 \\
0 & 1 & 0 & 0 & 1 & 0 \\
0 & 1 & 1 & 0 & 1 & 1 \\
1 & 0 & 0 & 1 & 0 & 0 \\
1 & 0 & 1 & 1 & 0 & 1 \\
1 & 1 & 0 & 1 & 1 & 1 \\
1 & 1 & 1 & 1 & 1 & 0 \\
\hline
\end{tabular}

\bigskip

The above truth table indicates that the input-output relation of the T-F gate 
is such that
$a'=a$, $b'=b$, $c'= c \oplus a \cdot b$ (where $\oplus$ denotes the exclusive 
OR Boolean function and the ``dot-product'' denotes the Boolean AND function). 
The variables $a$ and $b$ are called 
control bits and the variable $c$ is the target bit. The control bits 
pass through the gate without change. The target bit is flipped only if
both control bits are logical 1.

We wrote the A-matrix of the basic 1-bit inverter in Equation \ref{A-matrix}
using the computational basis states $|0>$ and $|1>$. We can write an 
A-matrix for the 3-bit T-F gate using the 3-bit computational basis states 
$|000>$,
$|001>$, $|010>$, $|011>$, $|100>$, $|101>$, $|110>$, and $|111>$ (here the 
basis states are possible values of $|abc>$):

\begin{eqnarray}
A_{Toffoli-Fredkin}
 & = &
 \left[ \begin{array}{cccccccc}
             1 & 0 & 0 & 0 & 0 & 0 & 0 & 0 \\
             0 & 1 & 0 & 0 & 0 & 0 & 0 & 0 \\
             0 & 0 & 1 & 0 & 0 & 0 & 0 & 0 \\
             0 & 0 & 0 & 1 & 0 & 0 & 0 & 0 \\
             0 & 0 & 0 & 0 & 1 & 0 & 0 & 0 \\
             0 & 0 & 0 & 0 & 0 & 1 & 0 & 0 \\
             0 & 0 & 0 & 0 & 0 & 0 & 0 & 1 \\
             0 & 0 & 0 & 0 & 0 & 0 & 1 & 0 \\
             
\end{array}   \right]    
\label{TF}
\end{eqnarray}

The A-matrix is not diagonal, but it is obviously unitary. The unitarity 
guarantees reversibility.

Lloyd has proposed an experimental realization of the T-F gate \cite{lloyd}.
Consider three bistable atoms $A$, $B$ and $C$. The excited states encode
the binary bit 1 and the ground states the binary bit zero. The energy 
separation between the excited and ground state of atom $C$ depends on
whether the atoms $A$ and $B$ are both excited or not. An electromagnetic
pulse (called a $\pi$-pulse) is applied to the 3-atom system which is 
resonant with the energy separation between the excited and ground states
of atom $C$ when both $A$ and $B$ are in excited states. This pulse
will flip atom $C$ from excited to ground state (or vice versa), thereby
realizing the operation of a T-F gate. A quantum dot realization of a 
T-F gate, based on an exact calculation of the dependence of the energy
separation between the excited and ground states of $C$ on the states of $A$ and 
$B$ has been reported by us \cite{bandy2}.
 
There is a 
vast body of work dedicated to different implementations of reversible 
(dissipationless) gates using various physical systems. Some are mechanical
in nature
such as billiard balls, some are electrical, some optical and yet some others
are chemical.  The first 
generation of these ``dissipationless'' systems comprised logically reversible 
gates that dissipate less than $kTln2$ amount of energy (but not necessarily 
zero energy) per bit flip.  Concrete proposals for such systems were advanced by
Bennett \cite{bennet1,bennet2}, Toffoli \cite{toffoli}, Fredkin \cite{fredkin}, 
Likharev \cite{likharev} and Landauer \cite{landauer2} among others.
None of these proposals envisioned nanoelectronic (or even semiconductor 
solid-state) implementation. Only recently, nanoelectronic versions based on 
single electron parametron
\cite{likharev_science},  two-level atoms or molecules \cite{lloyd}, and  
Coulomb or exchange-coupled quantum dots \cite{bandy1, bandy2} have appeared in
the literature.

\section{Quantum gates}

Quantum gates are of course always reversible in nature since they rely on
the quantum mechanical evolution of a system to operate on qubits.
The quantum mechanical evolution is described by the Schr\"odinger 
equation which obeys time reversal symmetry. All quantum mechanical 
evolutions are unitary in that the wave function $\psi(t)$, at any time
$t$, is related to the wave function $\psi(0)$ at time $t$ = 0, by the 
relation
\begin{equation}
\psi(t) = e^{-i H t/\hbar} \psi(0)
\end{equation}
where $H$ is the Hamiltonian of the system and a Hermitean operator.
Therefore the operator $e^{-i H t/\hbar}$ is unitary.

\subsection{The strange nature of quantum gates}

Quantum gates are strange entities that may not have any classical 
analog. Let us say that we want to invert a classical bit by operating
with an ordinary  $NOT$ gate. We choose to do this by operating on the input bit 
with two strange
gates in succession that we call ``square-root-of-NOT'' gates denoted by 
$\sqrt{NOT}$. The name comes about from the fact that the 
successive operation of two gates can be represented mathematically 
as the scalar product of their A-matrices. Thus, the logic operation
that we are after can be written as 
\begin{equation}
\sqrt{NOT} \cdot \sqrt{NOT} \equiv NOT 
\label{NOT}
\end{equation}

What makes this gate ``quantum'' (or at least non-classical) is that it
is impossible to have a single-input/single-output classical binary gate that
works in this fashion \cite{collins}. If the $\sqrt{NOT}$ gate were classical,
it should have an output of 0 or 1 for each possible input of 0 or 1. Suppose
that we define the action of a $\sqrt{NOT}$ gate as the pair of transformations
\begin{eqnarray}
\sqrt{NOT}_{classical} (0) = 0 \nonumber \\
\sqrt{NOT}_{classical} (1) = 0
\end{eqnarray}
Then, two consecutive applications of this gate will invert a 1 successfully,
but not a 0, thereby violating Equation \ref{NOT}. We can try any other 
transformation:
\begin{eqnarray}
\sqrt{NOT}_{classical} (0) = 0 \nonumber \\
\sqrt{NOT}_{classical} (1) = 1
\end{eqnarray}
This particular transformation does not invert any bit at all. In fact, it
is impossible to define $\sqrt{NOT}$ classically such that two successive
operations of this gate will reproduce the behavior of a $NOT$ gate.

The A-matrix of a $\sqrt{NOT}$ gate can be defined as 
\begin{eqnarray}
A_{\sqrt{NOT}}
 & = &
 \left[ \begin{array}{cc}
             {{1 + i}\over{2}} & {{1 - i}\over{2}} \\
             {{1 - i}\over{2}} & {{1 + i}\over{2}} \\
             \end{array}   \right]    
\label{sqNOT}
\end{eqnarray}
Note that 
\begin{eqnarray}
\left[ \begin{array}{cc}
             {{1 + i}\over{2}} & {{1 - i}\over{2}} \\
             {{1 - i}\over{2}} & {{1 + i}\over{2}} \\
             \end{array}   \right]
             \cdot 
             \left[ \begin{array}{cc}
             {{1 + i}\over{2}} & {{1 - i}\over{2}} \\
             {{1 - i}\over{2}} & {{1 + i}\over{2}} \\
             \end{array}   \right]
             =
             \left[ \begin{array}{cc}
             0 & 1 \\
             1 & 0 \\
             \end{array}   \right]
\end{eqnarray}
so that $A_{\sqrt{NOT}} \cdot A_{\sqrt{NOT}}$ = $A_{NOT}$, as required. 
Moreover, the 
matrix $A_{\sqrt{NOT}}$ is unitary which it must be.

Note that if a $A_{\sqrt{NOT}}$ gate is fed an input data string (0,1),
it produces an output data string ($\alpha$, $\beta$) given by
\begin{eqnarray}
\alpha = {{1 + i}\over{2}}|0> + {{1 - i}\over{2}}|1> \nonumber \\
\beta = {{1 - i}\over{2}}|0> + {{1 + i}\over{2}}|1>
\end{eqnarray}
where the output bits are coherent superpositions of $|0>$ and $|1>$ which 
are qubits!. Thus, the $A_{\sqrt{NOT}}$ gate is a quantum gate.

The A-matrix in equation (1) is not the only A-matrix that can represent 
a $\sqrt{NOT}$ gate. Consider the A-matrix
\begin{eqnarray}
A_{\sqrt{NOT'}}
 & = &
 {{1}\over{\sqrt{2}}}\left[ \begin{array}{cc}
             1 & 1 \\
             -1 & 1\\
             \end{array}   \right]    
\label{sqNOT'}
\end{eqnarray}

This matrix is also a valid A-matrix for a $\sqrt{NOT}$ gate since 
its square is the A-matrix of the $NOT$ gate (never mind about the negative
sign that you get when you operate the A-matrix twice on the data string
(0,1). In binary Boolean algebra, we will make no distinction between
1 and -1; the sign is unimportant). If this gate is fed 
an input data string (0,1),
it produces an output data string ($\alpha'$, $\beta'$) given by
\begin{eqnarray}
\alpha ' = {{1}\over{\sqrt{2}}}|1> + {{1}\over{\sqrt{2}}}|0> \nonumber \\
\beta ' = {{1}\over{\sqrt{2}}}|1> - {{1}\over{\sqrt{2}}}|0>
\end{eqnarray}
which are also qubits (no surprise there).

Now a measurement of the output (either $|\alpha'>$ or $|\beta'>$) yields the 
answer 0 (state $|0>$) or 1 (state $|1>$) with equal probability 1/2, yielding
a {\it perfectly random} generator of bits 0s and 1s. A single computation
with a single gate yields a perfectly random bit! This is an extension
of ``quantum parallelism''. First, a classical algorithm will require 
more than one gate and many computational steps to generate a
random number. More importantly, the number will be {\it never truly random}.
Mathematically, there is {\it no function} that generates a true random
number (only pseudo-random numbers can be generated using algorithms).
Thus, a classical Turing machine can only generate pseudo random numbers;
it cannot generate a true random number. This also shows that the classical
Turing machine is not really universal since it cannot model the quantum 
mechanical process of generating a true random number.

\subsection{Universal quantum gates}

The first universal quantum gate that was shown to be universal was not
a 2-qubit gate, but rather a 3-qubit gate due to Deutsch. It is fashioned 
after the universal classical gate of Toffoli and Fredkin. Using the 
computational basis states $|000>$,
$|001>$, $|010>$, $|011>$, $|100>$, $|101>$, $|110>$, and $|111>$, the A-matrix
of the Deutsch gate is given by

\begin{eqnarray}
A_{Deutsch}
 & = &
 \left[ \begin{array}{cccccccc}
             1 & 0 & 0 & 0 & 0 & 0 & 0 & 0 \\
             0 & 1 & 0 & 0 & 0 & 0 & 0 & 0 \\
             0 & 0 & 1 & 0 & 0 & 0 & 0 & 0 \\
             0 & 0 & 0 & 1 & 0 & 0 & 0 & 0 \\
             0 & 0 & 0 & 0 & 1 & 0 & 0 & 0 \\
             0 & 0 & 0 & 0 & 0 & 1 & 0 & 0 \\
             0 & 0 & 0 & 0 & 0 & 0 & A & B \\
             0 & 0 & 0 & 0 & 0 & 0 & B & A \\
             
\end{array}   \right]    
\label{TF1}
\end{eqnarray} 
where 
\begin{equation}
A = i e^{{{i \pi \alpha}\over{2}}} ( 1 + e^{i \pi \alpha} ) ~ and ~
B = i e^{{{i \pi \alpha}\over{2}}} ( 1 - e^{i \pi \alpha} )
\label{AB}
\end{equation}
and $\alpha$ may be an irrational number.

It is easy to check that $A_{Deutsch}$ is unitary. The T-F gate can be 
synthesized by connecting a number of Deutsch gates in series. The number of
Deutsch gates required for this is $N$ where $N$ must satisfy the relation
$[A_{Deutsch}]^N$ = $A_{Toffoli-Fredkin}$. It is always possible to find
an integer value of $N$ as long as $\alpha$ is an irrational number.
Noting that
\begin{eqnarray}
\left[ \begin{array}{cc}
       i e^{{{i \pi \alpha}\over{2}}} ( 1 + e^{i \pi \alpha} ) &      
       i e^{{{i \pi \alpha}\over{2}}} ( 1 - e^{i \pi \alpha} ) \\
       i e^{{{i \pi \alpha}\over{2}}} ( 1 - e^{i \pi \alpha} ) &
       i e^{{{i \pi \alpha}\over{2}}} ( 1 + e^{i \pi \alpha} ) \\      
\end{array}   \right]^N    
=
i^N e^{{{i N \pi \alpha}\over{2}}}
\left[ \begin{array}{cc}
       i e^{{{i \pi \alpha}\over{2}}} ( 1 + e^{i \pi \alpha} ) &      
       i e^{{{i \pi \alpha}\over{2}}} ( 1 - e^{i \pi \alpha} ) \\
       i e^{{{i \pi \alpha}\over{2}}} ( 1 - e^{i \pi \alpha} ) &
       i e^{{{i \pi \alpha}\over{2}}} ( 1 + e^{i \pi \alpha} ) \\      
\end{array}   \right]
\end{eqnarray}
the quantity $e^{i N \pi \alpha}$ can be made arbitrarily close to any complex 
number of unit norm by changing $N$. Choosing $N$ such that $e^{i N \pi \alpha}$ 
= -1 yields the Toffoli-Fredkin gate \cite{vwani2}.

\subsection{2-qubit universal quantum gates}

Recently, it has been shown that there exist 2-qubit quantum gates
which are universal. The proof of universality of these gates was given by
DiVincenzo \cite{divincenzo1} using Lie algebra; however, it can also be 
shown that the Deutsch gate can be realized with these 2-qubit gates. This, in 
itself,
is sufficient proof of universality. The Deutsch gate is not the 
most primitive universal quantum gate, the DiVincenzo gate is. Later, it was
shown by Seth Lloyd that almost any 2-qubit gate is universal \cite{lloyd1}.

The computational basis of the 2-terminal (2 inputs and 2 outputs) universal
quantum gate can be chosen as $|00>$, $|01>$, $|10>$, $|11>$. In this basis,
the A-matrix of the universal 2-qubit gate is
\begin{eqnarray}
A_{DiVincenzo}
 & = &
 \left[ \begin{array}{cccc}
             1 & 0 & 0 & 0 \\
             0 & 1 & 0 & 0  \\
             0 & 0 & A & B  \\
             0 & 0 & B & A  \\
\end{array}   \right]    
\label{DiVincenzo}
\end{eqnarray} 
where $A$ and $B$ have been given by Equation \ref{AB}.

\section{Solid State Realizations of Quantum Gates}

In 1996, this author and a co-worker proposed a simple 
spin based quantum inverter utilizing two exchange coupled
quantum dots \cite{bandy1}. It is not a universal 
gate, but it relies on quantum mechanics to elicit the 
Boolean logic NOT function and is described below.

\begin{figure}
\epsfxsize=3.4in
\centerline{\epsffile{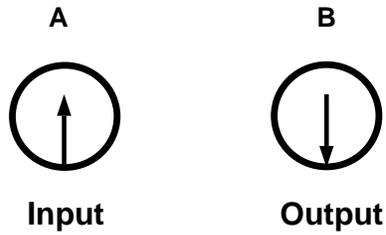}}
\caption{\tt Two exchange-coupled electrons in two quantum dots $A$ and $B$. The 
ground state of this system is anti-ferromagnetic; 
 the spin orientation of one electron is anti-parallel to that 
of the other. If spin is used to encode logic bits, then this system
acts as an inverter if we consider one quantum dot ($A$) as the input
port and the other ($B$) as the output port. The spin orientation in
$A$ can be aligned by a SPSTM tip thus enabling a ``write'' operation.
The spin orientation in $B$ can be read by a SPSTM tip thus enabling 
a ``read'' operation. The same system can act as a quantum inverter
since the ground state wave function is an entangled state and each quantum
dot can exist in a coherent superposition of ``upspin'' and ``downspin'' 
states. Reproduced from ref. \cite{bandy1} with permission of Academic Press. 
\label{spinNOT}}
\end{figure}

Consider two single electrons housed in closely spaced quantum dots as
shown in Fig. \ref{spinNOT}.  
We will assume that there is only one size quantized level in each dot. A
weak magnetic field $H_z$ is applied globally to define a spin polarization
direction. The ground state of
this two-electron system is antiferromagnetic \cite{ashcroft} with the two spins 
antiparallel
(one spin will be aligned along the magnetic field and the other opposite
to it).
If  spin polarization is used to encode binary bits such that up-spin
represents binary bit 1 and down-spin binary bit 0, then the spin polarization 
of one electron is the logic complement of that of the other when the system
is in ground state. This can be the
basis of an inverter. We can orient the spin polarization in one dot (with 
a local magnetic field)
to conform to the input bit and the output will automatically be the 
inverse of the input.

Let us study this system in a little more detail.
Even a 2-electron system is quite complicated really when it comes to quantum 
mechanics.
Any quantum system is described by a Hamiltonian which sums up the 
energy contributions to the electrons.
The so-called Hubbard Hamiltonian
for this system is

\pagebreak

\begin{eqnarray}
\cal H & = & \sum_{i \sigma} \left [ \epsilon_0 n_{i \sigma} + g \mu_B H_i 
sign(\sigma) \right ] + \sum_{<ij>} t_{ij} \left [ c^{+}_{i \sigma}c_{j \sigma} 
+
 h.c \right ] + \sum_{i}U_i n_{i \uparrow} n_{i \downarrow} \nonumber \\
 & & +  \sum_{<ij> \alpha \beta} J_{ij} c^{+}_{i \alpha} c_{i \beta} c^{+}_{j 
\beta}
c_{j \alpha} + H_z \sum_{i \sigma} g \mu_B n_{i \sigma} sign (\sigma)
\end{eqnarray}
where the first term denotes the electron energy in the $i$th dot, ($H_i$ is
a z-directed magnetic field applied selectively to the $i$th dot with, say,
a spin polarized scanning tunneling microscope (SPSTM) tip, to orient its spin 
polarization), the second
term denotes the hopping between the dots, the third term is the Coulomb 
repulsion within the $i$th dot, the fourth term is the exchange interaction
between nearest neighbor dots, and the last term is the Zeeman splitting
induced by the globally applied magnetic field $H_z$ directed along the
z-direction.

Molotkov and Nazin \cite{molotkov} have simplified this Hamiltonian to the 
Heisenberg model
which yields
\begin{equation}
{\cal H} = J \sum_{<ij>} \sigma_{zi} \sigma_{zj} + J \sum_{<ij>} \left [ 
\sigma_{xi} \sigma_{xj} + \sigma_{yi} \sigma_{yj} \right ] + \sum_{input~dots} 
\sigma_{zi}
h_{zi}^{input} ~~~~~~~~~~ (J>0)
\end{equation}
where we have neglected the global magnetic field $H_z$. The quantity $J$
is the exchange splitting and $h_{zi}^{input}$ is the Zeeman splitting
caused by the local magnetic field applied with, say  an SPSTM tip,
 to the $i$th dot to orient the
spin(s) of its electron(s). 

The above Hamiltonians describe any number of dots, each containing
any number of electrons. Here, we are concerned with the special case of
 just two dots each containing only one electron. 
We will call these two dots $A$ and $B$, where $A$ is the input dot
(whose spin polarization is set by an external SPSTM tip to conform to the input 
bit) and $B$ is the output dot.
Let us consider the case when the 
input bit corresponds to ``upspin''.

In the basis of two electron states, the Hamiltonian
in Equation (2) can be written as

\begin{eqnarray}
\begin{array}{cccc}
 |\downarrow \downarrow> & | \downarrow \uparrow > & |\uparrow \downarrow> 
 &
 |\uparrow \uparrow> 
\end{array} \nonumber \\
 \left( \begin{array}{cccc}
             h_A + J & 0 & 0 & 0 \\
             0 & h_A - J & 2J & 0 \\
             0 & 2J & - h_A - J & 0 \\
	     0 & 0 & 0 & -h_A + J \\
\end{array}   \right)  &  
  \begin{array}{c}
              |\downarrow \downarrow> \\
              | \downarrow \uparrow > \\
              |\uparrow \downarrow> \\
	      |\uparrow \uparrow>\\
\end{array}
\end{eqnarray}
where $h_A$ is the Zeeman splitting caused by the externally applied local 
magnetic field in input dot A.

\bigskip

The eigenenergies and eigenstates of the above Hamiltonian are  found by
diagonalizing

\bigskip

\begin{tabular}{|l|l|}
\hline
Eigenenergies & Eigenstates \\
\hline
$h_A + J$ & $|\downarrow \downarrow>$ \\
$-J + \sqrt{h_{A}^{2} + 4J^{2}}$ & $\sqrt{{{1}\over{2}} \left (1 + 
{{h_A}\over{\sqrt{h_A^{2} + 4J^{2}}}} \right )}  |\uparrow \downarrow> + 
\sqrt{{{1}\over{2}} \left (1 - {{h_A}\over{\sqrt{h_A^{2} + 4J^{2}}}} \right ) } 
|\downarrow \uparrow>$ \\
$-J - \sqrt{h_{A}^{2} + 4J^{2}}$ & $\sqrt{{{1}\over{2}} \left (1 - 
{{h_A}\over{\sqrt{h_A^{2} + 4J^{2}}}} \right ) } |\uparrow \downarrow> - 
\sqrt{{{1}\over{2}} \left (1 + {{h_A}\over{\sqrt{h_A^{2} + 4J^{2}}}} \right ) } 
|\downarrow \uparrow>$ \\ 
$-h_A + J$ & $|\uparrow \uparrow>$ \\
\hline
\end{tabular}

\bigskip

In the absence of any applied local magnetic field ($h_A$ = 0), the ground
state energy is -3$J$ and the ground state wave function is 
${{1}\over{\sqrt{2}}} (
|\uparrow \downarrow> - |\downarrow \uparrow> )$ which is an entangled 
state in that neither the input dot nor the output dot has a definite
spin polarization.

In ref. \cite{bandy1}, we showed that if the system is initially  in the 
ground state and a local magnetic field is applied to the input dot A
 at time $t$=0 to align its spin polarization to the ``up'' state, then unitary 
evolution of the system according to 
\begin{equation}
\psi (t) = exp[ -i {\cal H} t/\hbar ] \psi(0),
\end{equation}
mandates that the wave function at time $t$, is given by 
\begin{equation}
\psi(t) = c_2(t) |\uparrow \downarrow > + c_3(t) |\downarrow \uparrow >
\end{equation}
where 
\begin{eqnarray}
c_2 (t) = {{e^{i J t/\hbar}}\over{\sqrt{2}}} \left [ cos (\omega t) 
- i \left ( {{h_A}\over{\hbar \omega}} + \sqrt{1 - {{h_{A}^{2}}\over{\hbar^2 
\omega^2}}} \right ) sin(\omega t) \right ] \nonumber \\
c_3 (t) = - {{e^{i J t/\hbar}}\over{\sqrt{2}}} \left [ cos (\omega t) 
- i \left ( {{h_A}\over{\hbar \omega}} - \sqrt{1 - {{h_{A}^{2}}\over{\hbar^2 
\omega^2}}} \right ) sin(\omega t) \right ]
\end{eqnarray}
and $\hbar \omega$ = $\sqrt{h_A^2 + 4 J^2}$

If the system were to act as an inverter, it should ultimately reach the 
state $| \uparrow \downarrow>$. This desired state however is not an eigenstate 
of the 
system. Consequently, the system will continue to evolve to a different state
unless a ``read'' operation collapses the wave function as soon as the 
desired state is reached. The time to reach the desired state, which 
we will designate the switching time, is given by
\begin{equation}
\tau_d = {{h}\over{4 \sqrt{h_{A}^{2} + 4 J^{2}}}} 
\end{equation}
Note that this time must be much shorter that $\hbar/kT$ in order to 
maintain reasonable coherence in quantum computing \cite{unruh}. This can be
achieved by making $J >> kT$. For 100 \AA~ diameter dots separated 
by 1 eV high and 20 \AA~ wide barriers, the exchange splitting $J$
can be on the order of 100 meV in semiconductors.

If the inverter's initial state is the ground state, then it is possible to 
switch the inverter {\it completely} 
(i.e. $c_2(\tau_d)$ = 1,  $c_1(\tau_d)$ = $c_3(\tau_d)$ = $c_4(\tau_d)$ = 0)
if $h_A$ = 2$J$. However, if the initial state is not the ground state,
then the inverter can never switch completely to the desired state, (i.e. 
$c_2(\tau_d)$ $<$ 1).
Nonetheless, we can still define a switching time as the delay 
that elapses before the closest approach to the desired state (in other words
the time required to reach the maximum value of $c_2$). This switching time is 
still given by Equation (6).
This equation also shows that the inverter will ``switch'' in a finite time 
{\it even if} the switching energy $h_A$ $\rightarrow$ 0. However, the 
``switching''  is ephemeral since the system will continue to evolve unitarily
unless a read operation collapses the wave function at the right juncture.

In this system, there is no dissipation whatsoever except during the 
read operation. Therefore, the product of dissipated energy and switching delay 
(necessary to complete the computation) can obviously be zero.
The product of applied energy and switching delay is
\begin{equation}
h_A \tau_d = {{h h_A}\over{4 \sqrt{h_{A}^{2} + 4 J^{2}}}}
\end{equation}

We immediately see that any energy-time uncertainty that might have been 
expected is violated
\begin{equation}
h_A \tau_d < {{\hbar}\over{2}} ~~~~~~if~ h_A < {{2J}\over{\sqrt{\pi^2 -1}}}
\end{equation}

Thus we can violate the uncertainty by making $h_A$ arbitrarily small
(and yet switching in a finite time).
It should be emphasized that $h_A$ is the energy applied to switch the 
inverter and is not necessarily {\it dissipated}. Even if it were 
completely dissipated, the above equation would still clearly show that 
there is 
no energy-time uncertainty limitation on (dissipated) energy-delay product, 
contrary to the popular
view espoused in \cite{bate,mead}. Landauer agrees that there is no
limit imposed by the uncertainty principle as per as dissipated energy 
is concerned \cite{landauer3}. In fact, concrete and detailed classical models 
of dissipationless 
computation have been provided by several authors  and numerous quantum 
mechanical 
models of dissipationless computation have also been forwarded starting with the
early work of Benioff \cite{benioff,benioff2}. These models require no 
dissipated energy,
but usually do require input energy to switch. What we see in the pathological
example above is that there is no energy-time uncertainty limit 
even when the energy concerned is the {\it applied energy} rather than the 
dissipated energy. Computation can proceed by applying arbitrarily small energy
to initiate the process.

\section{Coulomb Coupled Quantum Dots for Toffoli-Fredkin Gate and Quantum 
Computation}

In the previous section, we described a quantum inverter. An inverter however is 
not a universal gate. The Toffoli-Fredkin gate is a mathematically (and hence 
physically) reversible
three-bit {\it universal} gate with three inputs $A$, $B$, $C$ and three
outputs $A^{\prime}$, $B^{\prime}$ and $C^{\prime}$. 

As mentioned before, in this gate, $A^{\prime}$ = $A$ and $B^{\prime}$ = $B$. 
The bit $C^{\prime}$
= $\bar{C}$ only if $A$ = $B$ = 1. Otherwise, $C^{\prime}$ = $C$. 

A physical realization of the Toffoli-Fredkin gate that is most easily amenable
to nanoelectronic adaptations was proposed by Seth Lloyd \cite{lloyd} expanding 
on ideas set forth earlier by Mahler, et. al. \cite{mahler}. The Lloyd 
architecture consists of
an array of three weakly coupled quantum systems $A$, $B$ and $C$. Each system 
can exist in one of two energy states $E^{i}_{0}$ and $E^{i}_{1}$ (i $\in$ $A$, 
$B$, $C$) which represent
logic 0 and 1. Furthermore, $A$, $B$, $C$ are distinct systems such that the 
resonant energies $\hbar \omega_{i}$ = $E^{i}_{1} - E^{i}_{0}$ are different for 
each of them
($\omega_A \neq \omega_B \neq \omega_C$). A $\pi_i$
pulse is a radiation that obeys the condition
\begin{equation}
{{1}\over{\hbar}} \int {\vec \mu_B^{i}} \cdot {\hat e} {\cal E} (t) dt = \pi
\end{equation}
where ${\vec \mu_{B}^{i}}$ is the induced dipole moment between the ground state 
and excited state of the $i$-th system, ${\hat e}$ is the polarization 
unit vector of the incident radiation and ${\cal E}(t)$ is the magnitude of
the pulse envelope at time $t$. Such a pulse flips the $i$-th system from
the excited state to the ground state and vice-versa. It is assumed that 
the duration of this pulse is much shorter than the inverse of the 
spontaneous decay rate from the excited to the ground state.

Because of the nearest-neighbor interaction between $A$, $B$ and $C$, the 
resonant 
energies $\hbar \omega_i$ for each of them is no longer 
unique and depends on whether 
its nearest neighbors are in the excited or ground state.
Thus

\begin{eqnarray}
\omega_{A} & \rightarrow &  \omega_{0}^{A}, \omega_{1}^{A} \nonumber \\
\omega_{B} & \rightarrow & \omega_{00}^{B}, \omega_{01}^{B}, \omega_{10}^{B}, 
\omega_{11}^{B} \nonumber \\
\omega_{C} & \rightarrow & \omega_{0}^{C}, \omega_{1}^{C} 
\end{eqnarray}

Here, the subscripts on the right hand side refer to the states of the 
corresponding system's nearest neighbor(s). For instance, $\omega_{00}^{B}$ is 
the resonant 
frequency of system $B$ when its two neighbors $A$ and $C$ are both 
in their ground states. 

A Toffoli gate can be realized by shining a $\pi$ pulse with frequency
$\omega_{11}^{B}$. The state of $B$ is inverted only if both $A$ and 
$C$ are in their excited states. This characteristic realizes the truth table
of a Toffoli gate.

\subsection{Connecting Toffoli-Fredkin gates}

In order to do computation, one needs to connect various Toffoli gates.
This can be realized without physical wires if we have a linear array
consisting of units $ABC$, $ABC$, ... Computation is performed by
first initializing the array to the input with appropriate sequence
of $\pi$ pulses and then applying another series of $\pi$ pulses to 
complete the computation. This methodology was described in detail 
in Ref. \cite{lloyd}.

The above system can also be used to perform quantum computation if both
$\pi$ and $\pi/2$ pulses are used. A $\pi_i/2$ pulse puts the $i$-th
system in a state $1/\sqrt{2}(|1> - |0>)$ which is a ``qubit'' in the
coherent superposition of bits 1 and 2. By using an appropriate pulse
train, one can perform quantum computation.

Rolf Landauer, an ardent critic and astute examiner of all ``quantum schemes'',  
has criticized this implementation from two angles. First, this gate is not
truly dissipationless unless the $\pi$ or $\pi/2$ pulses can be recycled.
This would also require that they are not
distorted by interaction with the system. 
Second, interaction with the environment will cause errors and 
error correction will require dissipation. In principle, the 
latter objection is no longer serious in view of the recent 
advances in quantum error control coding \cite{Shor,calderbank,stean}. Errors 
can be
corrected by ``software'' rather than ``hardware'' countermeasures. Of course,
this is done at the expense of increased memory and a larger 
system may decohere more quickly than a smaller system. Therefore,  error 
correction comes with a cost, just as everything else does.

Landauer's first criticism is much more difficult to rebut. Photon recycling is 
not an unheard of concept in solid state systems 
\cite{melloch1,melloch2,melloch3,melloch4} but it is 
difficult. The requirement that 
the pulse shape remain undistorted in any wave guide is a tall order.
This would require the wave guide to have specific non-linearities 
so that the pulses essentially become solitons. At this time, one does not have 
a suitable design for such a recycler.

We will next examine a specific implementation
of Lloyd's generic ideas and provide a concrete example of a Toffoli-Fredkin
gate. This example is suitable for nanoelectronics and is due to  
Alexander Balandin and the author.

\subsection{Nanoelectronic version of a Toffoli-Fredkin gate} 

Consider an array of three quantum dots with  high barriers (Fig. 3).
Each houses a single conduction band electron.

\begin{figure}
\epsfxsize=4.2in
\centerline{\epsffile{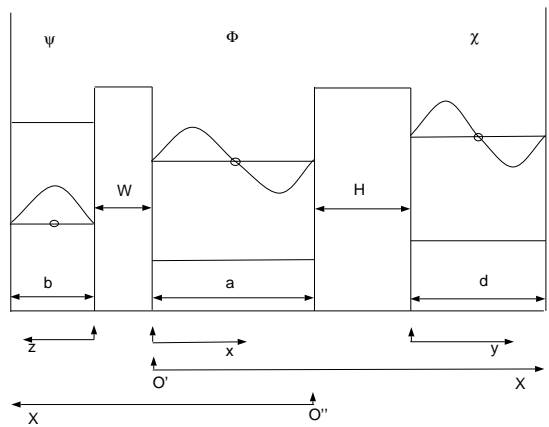}}
\caption{\tt The potential profile for a three-dot system showing the wave 
function 
envelopes for the ground and excited states. Reproduced from ref. \cite{bandy2}
with permission of Academic Press.
\label{TFgate}}
\end{figure}

For high enough barriers, we can neglect any overlap between the wave functions 
of electrons
in adjacent dots, and write
the many-body wave function of the three-electron system as a product of three 
single particle 
wave functions in the Hartree approximation
\begin{equation}
\Psi_{n,k,l}\equiv \Psi_{n,k,l}(x_{1},x_{2},x_{3})=
\psi_{n}(x_{1})\phi_{k}(x_{2})\chi_{l}(x_{3}),
\label{wave function}
\end{equation}
where $\psi_{n}(x_{1})$, $\phi_{k}(x_{2})$, and $\chi_{l}(x_{3})$ are the single 
electron
envelope functions for the first, second and third dots, respectively. 
Subscripts $n,~k,~l$
denote conduction subband levels. We assume that each dot has  two bound states
in the conduction band so that each electron can
occupy either ground state (n=1, k=1, or l=1) or the excited state (n=2, k=2, or 
l=2). These two states encode logic bits 1 and 0.

Owing to Coulomb interaction, the resonant frequency for transitions between the 
excited and ground states in one dot depends on whether the electron(s) in the 
neighboring dot(s) are in the excited or ground state. For the central dot,
this means that
$\omega_{11} \neq \omega_{10} \neq \omega_{01} \neq \omega_{00}$. This forms the 
basis
of a Toffoli-Fredkin gate. Note that in reality, a single gate only
requires that $\omega_{11}$ be distinct. However, arbitrary data 
manipulation requires that all $\omega$'s be distinct.

Let us calculate the resonant transition frequency of the 
central quantum well as a function of two adjacent wells' states. We denote the 
widths of
the well as $d,$ $a,$ and $b$ (see Fig. 3) and will refer to them as the 
``left'' well (L),
the ``central'' well (C), and the ``right'' well (R), respectively. 
The barier thicknesses are $W$ and $H$. 
The first order perturbation corrections to the energy of 
the $k$ subband of the C well are given by the expression
\begin{equation}
E_{n,k,l}=E_{k}+ <\Psi_{n,k,l}|V(x_{3}-x_{2})|\Psi_{n,k,l}> +
<\Psi_{n,k,l}|V(x_{1}-x_{2})|\Psi_{n,k,l}>,
\label{energy}
\end{equation}
where $V(x_{i}-x_{j})$ is the Coulomb interaction terms, $x_{i}$ are absolute 
coordinates
of electrons belonging to different wells, $m^{*}$ is the 
effective mass of the conduction band electron, $E_{k}$ is the unperturbed 
confined energy
which is given for the square well potential by a regular expression  
\begin{equation}
E_{k}=\frac{k^{2}\pi^{2}\hbar^{2}}{2m^{*}a^{2}}.
\end{equation}

To simplify the calculation, we define a set of local coordinate systems (see 
Fig. \ref{TFgate}).
Substituting in Eq. \ref{wave function}, the electron envelope functions for the 
square well 
potential
(written in local coordinates), 
 can be rewritten  as 
\begin{equation}
\Psi_{n,k,l}=\sqrt{\frac{2}{d}}sin(\frac{\pi n z}{d})
\sqrt{\frac{2}{a}}sin(\frac{\pi k x}{a}) \sqrt{\frac{2}{b}}sin(\frac{\pi l 
y}{b}).
\end{equation}
The distance between electrons in the R and C wells (in O' coordinate system) is
$x_{3}-x_{2}=a+W+y-x$, while the distance between electrons in the L and C 
wells (in O'' coordinate system) is
$x_{1}-x_{2}=a+H+z-x$. With the limits of integrations determined by the well 
boundaries, Eq. \ref{energy} now reads
\begin{equation}
E_{n,k,l}=\frac{\pi^{2}\hbar^{2}k^{2}}{2m^{*}a^{2}}+\frac{e^{2}}{\pi 
\epsilon^{*} ab}
\int_{x=0}^{a} \int_{y=0}^{b} \frac{sin^{2}(\pi kx/a) sin^{2}(\pi 
ly/b)}{a+W+y-x} dx dy
\end{equation}
\[
+\frac{e^{2}}{\pi \epsilon^{*} ad}
\int_{x=0}^{a} \int_{z=0}^{d} \frac{sin^{2}(\pi kx/a) sin^{2}(\pi 
nz/d)}{a+H+z-x} dx dz,
\]
where $\epsilon^{*}=f_{b} \epsilon_{b}+ f_{w} \epsilon_{w}$ 
is an effective dielectric constant of the system, $f_{b}$ ($f_{w}$) and 
$\epsilon_{b}$ ($\epsilon_{w}$) are the volume
fraction and dielectric constant of the barrier (well) material, respectively.
Note that with this definition $\epsilon^{*}$ is a function of well width
when the barrier thickness is fixed.
 
The energy of the transition between the first excited state ($k=2$) and 
the ground state ($k=1$) in the central well can now be written as a function
of the principal numbers $n$ and $l$ of the neighboring wells:
\begin{equation}
\Delta E_{n,l}=\frac{3\pi^{2}\hbar^{2}}{2m^{*}a^{2}}+
\frac{e^{2}}{\pi \epsilon^{*} ab}
\int_{x=0}^{a} \int_{y=0}^{b} \frac{sin(3\pi x/a) sin(\pi x/a) sin^{2}(\pi 
ly/b)}{a+W+y-x} dx dy
\label{7}
\end{equation}
\[
+\frac{e^{2}}{\pi \epsilon^{*} ad}
\int_{x=0}^{a} \int_{z=0}^{d} \frac{sin(3\pi kx/a) sin(\pi kx/a) sin^{2}(\pi 
nz/d)}{a+H+z-x} dx dz.
\]
Here we have utilized some trigonometrical equalities to simplify the result of 
the
subtraction $\Delta E_{n,l}=E_{n,2,l}-E_{n,1,l}$.

In order to be able to build a conditional quantum gate (or the Toffoli-Fredkin 
gate), the transition energy
$\Delta E_{n,l}$ should be different for all possible quantum state \{$|n>, 
|l>$\}:
$\Delta E_{1,2} \ne \Delta E_{2,1} \ne \Delta E_{1,1} \ne \Delta E_{2,2}.$
Obviously, the two states which are most
difficult to resolve are \{$|1>,|2>$\} and \{$|2>,|1>$\}. For convenience, 
we write here explicitly the energy difference between these two states
\begin{equation}
\Delta E_{1,2}-\Delta E_{2,1}=
\frac{e^{2}}{\pi \epsilon^{*} a} 
\int_{x=0}^{a} 
\Bigl( \int_{y=0}^{b} \frac{sin(3\pi x/a) sin(\pi x/a) sin(3\pi y/b)
 sin(\pi y/b)}{b (a+W+y-x)} dy
 \label{17}
\end{equation}
\[
\int_{z=d}^{0} \frac{sin(3\pi kx/a) sin(\pi kx/a) sin(\pi z/d)sin(3\pi z/d)}{d 
(a+H+z-x)} dz
\Bigr) dx. \]
To derive the above equation, we used the fact that $sin^{2}(2\pi y/b)- 
sin^{2}(\pi y/b)=
sin(3\pi y/b) sin(\pi y/b)$, and changed the limits of integration. For the 
special case when $W=H$, this equation can be further simplified 
by substitution of variable in the integrand to
\begin{equation}
\Delta E_{1,2}-\Delta E_{2,1}=
\frac{e^{2}}{\pi \epsilon^{*} a} 
\int_{x=0}^{a} 
\int_{u=d}^{b} \frac{d sin(3\pi u/b) sin(\pi u/b)+b sin(3\pi u/d) sin(\pi u/d) 
 }{b d (a+H+u-x)} 
 \label{18} 
\end{equation}
\[ \times  sin(3\pi kx/a) sin(\pi kx/a) du dx. \]
It is easy to see from the last equation that when the thicknesses of two 
peripheral wells and barriers are equal ($d=b$ and $W=H$), the states 
\{$|1>,|2>$\} and \{$|2>,|1>$\} are degenerate and
can not be resolved. This is a direct result of the symmetry of the system and 
can be
easily guessed without mathematical consideration. More interesting consequence 
of
Equations \ref{17} and \ref{18}  is that there exists a ratio of $b/d$ ($\neq$ 
1) such that the integral in Equation \ref{18}
vanishes, and the states are degenerate again. The physical
origin of this additional degeneracy will be discussed later.  
One should find  optimum values of well thicknesses $b$ and $d$
such that the states \{$|1>,|2>$\} and \{$|2>,|1>$\} are resolved. 

\subsection{Resonant energies in Coulomb coupled dots}

In calculating resonant energies in Coulomb coupled dots, we will concentrate 
mostly on two material systems. The first is $InAs$ characterized by light 
electron effective mass $m^{*} (InAs) = 0.023
m_{o}$ and strong dielectric screening $\epsilon (InAs) = 14.6$ ($m_{o}$ is
the free electron mass). The second is $CdS$ which is characterized by 
heavy electron effective mass $m^{*} (CdS) = 0.21
m_{o}$ and relatively weak dielectric screening $\epsilon (CdS) = 5.4$ for the
frequencies close to band gap resonance and  $\epsilon (CdS)$ approaching $3.1$
for the infrared region relevant to intraband transitions. 

In order to find optimum values of peripheral well thicknesses, 
we will calculate transition energy $\Delta E_{n,l}$ as a function of the R well
thickness $b$ while fixing the L well thickness $d$ and using it as a parameter. 
For simplicity, the thickness of the left barrier will be assumed to be equal to 
that on the right ($W=H$). These thicknesses will be assumed to be $20\AA$ 
unless otherwise stated. 
This value of the barrier thickness guarantees negligible barrier penetration
of the wave function. 

\begin{figure}
\epsfxsize4.2in
\centerline{\epsffile{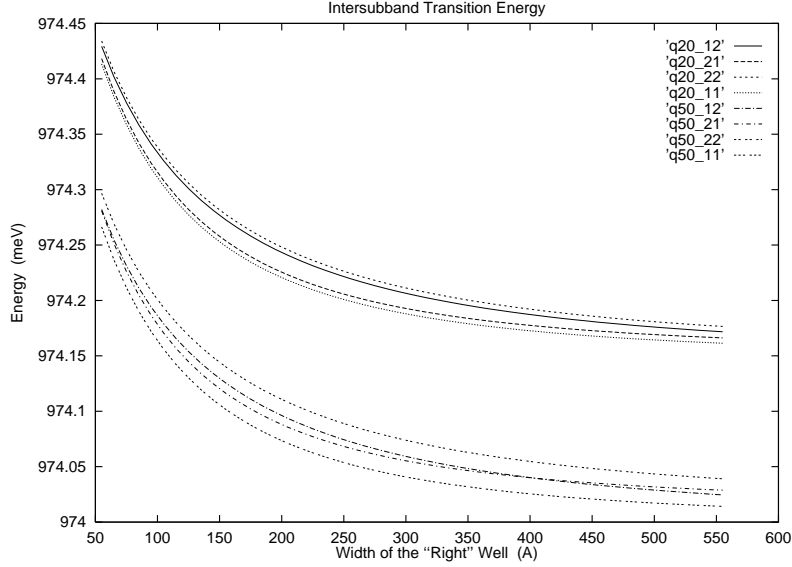}}
\caption{\tt Energy of the transition between the first excited and ground 
states
in the ``central well'' as a function of the width of the ``right'' well.
The upper family of curves corresponds to the ``left'' well width of 20\AA,~
the lower one corresponds to the ``left'' well width of 50\AA.~
The results are shown for all four possible combination of states in the two
extremal wells. Material parameters for InAs have been used. Reproduced from 
ref. \cite{bandy2}
with permission of Academic Press.
\label{transition}}
\end{figure}

In Fig. \ref{transition}, we present the energy of the transition between the 
first excited
state and the ground state in the C well as a function of the R well width 
($b$).
The curves are shown for $InAs$ quantum wells. The thickness of the C well is
$100\AA$.
The L well width is varied over two values: $20\AA$ and $50\AA$.
It is interesting to note that
the splittings between worst-resolved states attain maximum values
when $b$ is in the range $150-170\AA$. Moreover, the states  \{$|1>, |2>$\} and 
\{$|2>, |1>$\} are degenerate not only at $b=d=50\AA$ but also at $b \approx 
400\AA$.
At first, this may appear surprising.
The physical origin of the additional degeneracy lies in the fact
that Coulomb perturbation to the transition energies depends on the
distance between particles as well as on the electron envelopes which serve as
weight functions in the integrand in Eq. \ref{18}. Consequently, at some values 
of $b/d \ne 1$  
the integration over $u$ in Eq. \ref{18} vanishes. 

\begin{figure}
\epsfxsize4.2in
\centerline{\epsffile{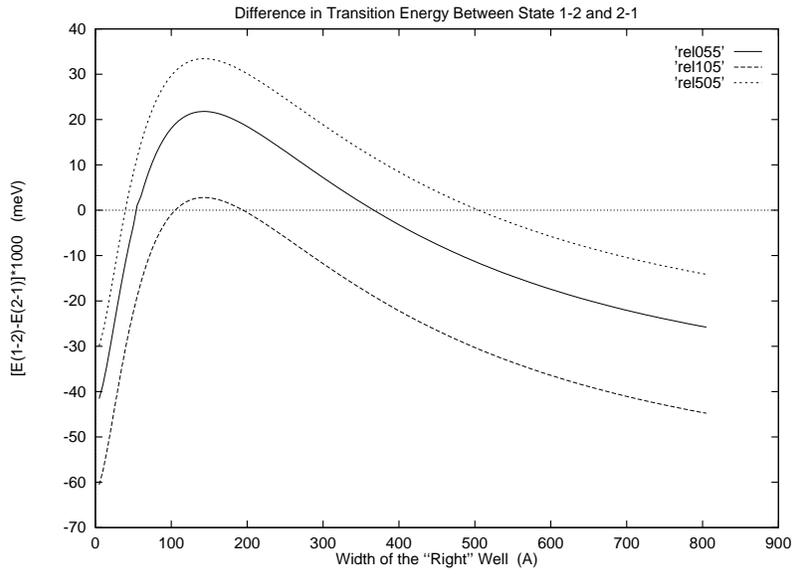}}
\caption{\tt The difference in transition energy $\Delta E_{1,2}-\Delta E_{2,1}$ 
as
a function of the width of the ``right'' well. The results are shown for
three different values of the width of the ``left'' well. The material
parameters used for calculation correspond to $InAs$. Reproduced from ref. 
\cite{bandy2}
with permission of Academic Press.
\label{worst-resolved_InAs}}
\end{figure}

\begin{figure}
\epsfxsize4.2in
\centerline{\epsffile{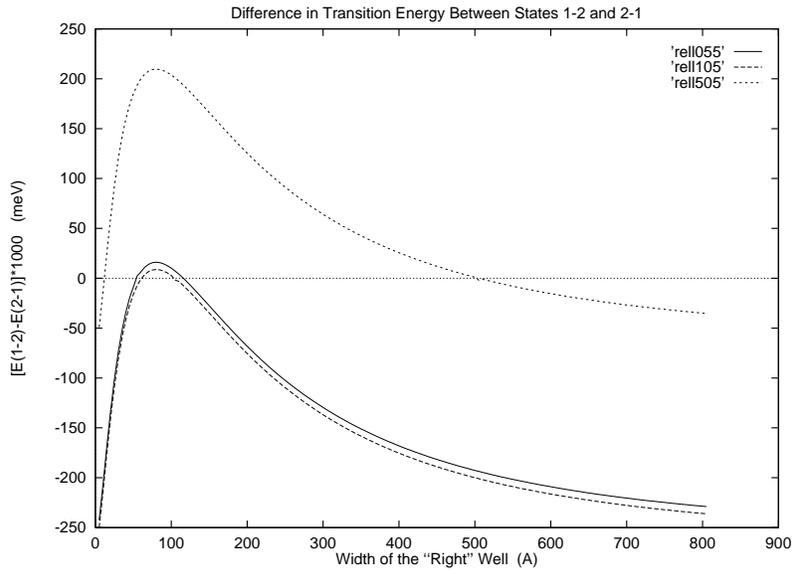}}
\caption{\tt The difference in transition energy $\Delta E_{1,2}-\Delta E_{2,1}$ 
as
a function of the width of the ``right'' well. The results are shown for
three different values of the width of the ``left'' well. The material
parameters used for this calculation correspond to $CdS$. Reproduced from ref. 
\cite{bandy2}
with permission of Academic Press.
\label{worst-resolved_CdS}}
\end{figure}

In order to examine the behavior of the worst resolved states  \{$|1>, |2>$\} 
and 
\{$|2>, |1>$\}, we plot separately the difference in transition energy between
these two states as a function of the R well thickness (see Fig. 
\ref{worst-resolved_InAs}). 
The L well thickness is chosen to be $55\AA$ (solid line), $105\AA$ (dashed 
line), and
$505\AA$ (dotted line). As one can see, the average splitting is very small for
this system and represents a fraction of meV at its maximum. Each curve has an 
additional degeneracy ($\Delta E_{1,2}- \Delta E_{2,1}=0$) at $b/d \ne 1$ which 
should be avoided while
designing the logic gate. Since  $\Delta E_{1,2}-\Delta E_{2,1}$ does not
depend on $m^{*}$ (see Eq. \ref{7}) and strongly depends on $\epsilon^{*}$, one 
can  expect that the resolution of  the \{$|1>, |2>$\} and 
\{$|2>, |1>$\} states will be better for $CdS$ and any other system with
lower dielectric constant. This is indeed the case, and it is clearly 
seen in Fig. \ref{worst-resolved_CdS}. The splitting between states is an order 
of
magnitude higher for the $CdS$ system compared to the $InAs$ system.

It is intuitively clear that in order to increase the energy 
difference between \{$|1>, |2>$\} and
\{$|2>, |1>$\} states, one should introduce higher asymmetry into
the system. Apart from varying the R well thickness, it is possible
to break the symmetry by changing the barrier thickness.
One can notice that increasing the width of one of the barriers 
decreases the energy separation between  \{$|1>, |1>$\} and
\{$|2>, |2>$\} states but increases it for 
\{$|1>, |2>$\} and \{$|2>, |1>$\} states. The
degeneracies ($b=50\AA$ and $b\approx 400\AA$), 
 for the uniform barrier system,
disappear for a system with different barriers. This is of course 
also desirable from architectural considerations. 
Another potential way of increasing the energy splitting 
between \{$|1>, |2>$\} and \{$|2>, |1>$\} states is through engineering
the potential profile. One well can be rectangular and the other 
parabolic. This may be achieved through dopant grading.

\section{A 2-qubit ``spintronic'' universal quantum gate}

Today, most experimental effort in realizing semiconductor solid-state 
versions of universal 2-qubit gates is focused on the realization of a
quantum controlled NOT or XOR gate. These are the two most popular gates.
In these gates, one of the two qubits is called the 
target qubit and the other is the control qubit. The basic operation of 
the controlled NOT gate has two ingredients: (i) arbitrary rotation of 
the target qubit, and (ii) controlled rotation of the target qubit
through specified angles if and only if the control qubit is in a 
certain state. The controlled XOR gate has  similar dynamics: (i) arbitrary 
rotation of 
the target qubit, and (ii) a so-called ``square-root-of-swap'' operation
where the quantum states of the target and control qubits are half-way
interchanged.  More on this operation later.

There is now a consensus that in semiconductor solid state realizations,
it is advantageous to encode the qubit in spin degrees of freedom of an electron 
or hole
rather than charge degrees of freedom  
because of the much longer 
coherence times associated with spin. The charge coherence time 
saturates to about 1 nsec in most solids as the temperature is
lowered to microkelvins. This is supposedly due to coupling of
the electron to the zero-point motion of phonons (quantum noise) 
\cite{jariwalla}.
In contrast, the spin coherence time of an electron can be as long as a few 
milliseconds
in materials such as silicon where the Land\'e g-factor is close to 2
(the free electron's g-factor) \cite{faher}. Nuclear spins can live even longer, 
perhaps an hour. Compound semiconductors may exhibit somewhat shorter 
electron spin coherence times, but spin coherence times 
as long as 100 ns have been experimentally demonstrated in
n-type GaAs at the relatively balmy temperature of 5 K \cite{kikkawa}. 
Consequently, there have been a number
of ``spintronic'' proposals for quantum gates.

\subsection{The Kane model}

The first comprehensive solid state scalable realization of a universal
2-qubit spintronic quantum gate is due to Kane \cite{kane}. Qubits are encoded 
in the nuclear spin orientations of $^{31}$P dopant atoms in silicon. These
nuclear spins have very long coherence times, possibly several milliseconds
at 4.2 K temperature. A particular target qubit (nuclear spin) is rotated 
by bringing the spin splitting energy of that nucleus in resonance 
with a global ac magnetic field for specific duration. The spin-splitting 
energy is altered (using the hyperfine interaction)
by changing the wave function of the electron bound to the target nucleus. This 
is done by
applying a potential to that electron via a lithographically defined 
gate (called an ``A-gate'') placed precisely on top of that nucleus. 
The gate potential attracts or repels the electron thereby changing its 
wave function, and hence the nucleus' spin-splitting energy. The 2-qubit 
operation 
is achieved by yet another set of lithographically defined gates (called 
``J gates'') which raise or lower a potential barrier between two neighboring
nuclei's electrons (one nucleus is the ``target'' and the other is the 
``control'' qubit). As the barrier is lowered, the wave functions of the two 
adjacent electrons
overlap and the resulting exchange interaction lowers the energy of the 
singlet state with respect to the triplet state (as long as there is 
no magnetic field to induce a Zeeman splitting that exceeds the exchange 
splitting). Thus, the rotation of the target qubit by the global ac
magnetic field can be made conditional on the spin state of the control 
qubit. This realizes the quantum controlled NOT operation.

A closely related proposal that hosts qubits in electronic spin,
rather than nuclear spin, has been forwarded by Vrijen et. al. \cite{vrijen}.
This model eliminates the need to transduce the qubit between nuclear and 
electron spin. The penalty is the 
Instead of using hyperfine interaction to change the spin splitting 
energy in a target qubit, they utilze a Si/SiGe hertostructure. The 
$^{31}$P dopant atom has to be placed exactly at the interface of 
this two materials and a lithographically delineated gate has to
be aligned exactly on top of this buried dopant atom. The lateral 
alignment may be achieved by ion-implanting the dopant through a 
mask \cite{mckinnon}, but the vertical alignment is much more difficult.
One has to virtually eliminate ``straggle'' in ion implantation. This is 
a vey tall order.

An additional complication in these models is the read/write mechanism
for qubits. Reading is performed by ``measuring'' the spin polarization 
electronically
without the aid of anything analogous to a Stern-Gerlach apparatus. 
One suggested possibility is to use Paul principle \cite{vrijen}.
An electron is maintained in a quantum dot in given spin state. The electron
to be probed is injected towards that quantum dot. If the two spins are
anti-parallel, there is no Pauli blockade (in addition to the normal
Coulomb blocakde), and the target electron will enter the dot. Therefore,
the dot's charge will change and this can be monitored by single-electron
electrometers. This is just the basic idea. Actual implementation may 
involve some refinements.

An earlier proposal, incorporating spintronics but without any mechanism 
for manuvering qubits, was presented by Privman \cite{privman}. This 
will not be discussed at any length since it does not constitute 
a true quantum gate.

\subsection{An alternate quantum dot based spintronic model}

We have proposed an alternate model that is much easier to synthesize
than the Kane model or its modification by Vrijen \cite{bandy}.
Particularly, the read/write scheme is vastly simpler since we propose
using ferromagnetic contacts for spin injection and detection. The notion
of using ferromagnetic contacts as spin polarizer and spin analyzer has 
been around for at least ten years \cite{das}. I describe this paradigm below.
 
\begin{figure}
\epsfxsize=5.8in
\epsfysize=3.4in
\centerline{\epsffile{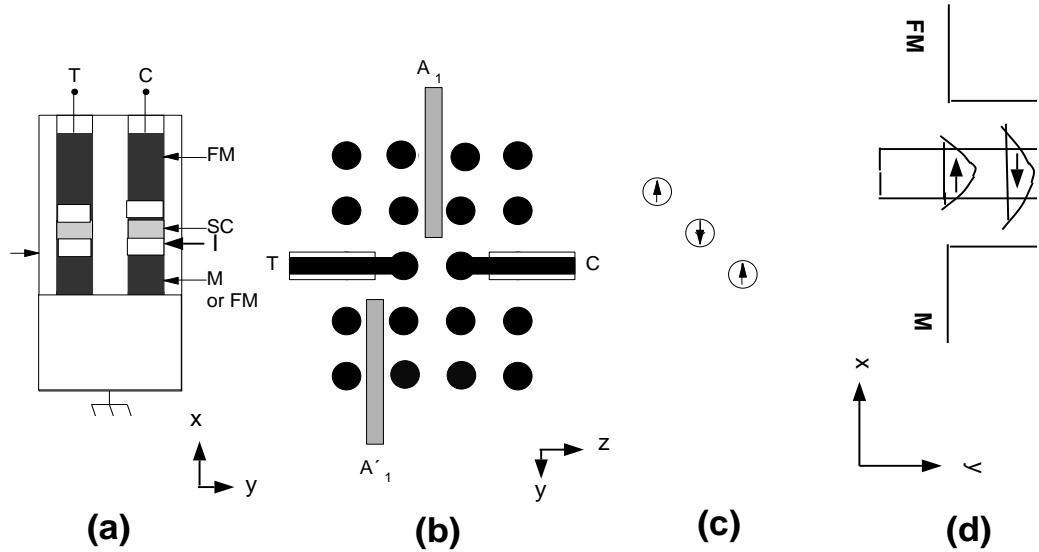}}
\caption[]{\tt Schematic diagram showing the
structure of a basic universal gate.
a) Cross section of a single gate.  {\sf FM}, {\sf M}, {\sf I} and {\sf SC} 
refer to
ferromagnetic, metallic, insulating  and semiconducting material respectively.
{\sf C} and {\sf T} are ohmic contacts to the control and target qubit
respectively.
b) Top view of a gate within the dot array.
Contacts labeled {\sf A} are for applying electrostatic potentials
to induce the Rashba effect, as well as turn on and off an 
exchange interaction between {\sf C} and {\sf T}.
c) A ``spin wire'' for interconnecting gates.
d) The conduction band diagram in the direction perpendicular to the 
heterostructures showing the spin-split levels. Spin splitting is 
caused by a combination of Zeemna effect and Rashba effect. For a 
``hole-qubit-structure'' (See Fig.. ), the valence band diagram can be obtained 
by inverting this diagram. Reproduced from ref. \cite{bandy_physica} with 
permission from Elsevier Science. \label{schematic}}
\end{figure}

Consider a semiconductor quantum dot capped by two spin-polarized ferromagnetic 
layers 
 (see Fig \ref{schematic}). 
Insulating layers are interposed between the semiconductor and 
ferromagnet to provide a confining potential for an electron 
in the semiconductor dot. A single spin-polarized quanton (electron or
hole) is injected into the semiconductor dot from a ferromagnetic contact, and 
trapped there by Coulomb blockade.

The ground state of the trapped quanton is automatically spin-split because of 
two effects: (i) the dc magnetic 
field caused by the ferromagnetic contacts introduces a Zeeman splitting,
and (ii) the intrinsic electric field at the interface between the 
quantum dot and the surrounding material causes an additional spin splitting  
because of the Rashba (spin-orbit)
interaction \cite{rashba,rossler,das}. The total spin splitting is a combination 
of the 
Rashba and Zeeman 
splittings.

The Rashba component can be varied by an external gate potential \cite{nitta}.
 Applying an electric field in the y-z plane using the contacts $A_1$, $A_1$'  
(see Fig. \ref{schematic}) modulates the energy splitting 
between the +x-polarized (spin-up) and -x-polarized (spin-down) 
states in the quantum dot. This controllability of the spin splitting 
energy with an external gate potential (because of the Rashba effect) is 
exploited to execute single qubit rotations
(rotating the spin through arbitrary angles). The  
dependence of the total spin splitting energy in this structure on the gate
potential is derived in the Appendix.

\subsection{Single qubit rotations}

To rotate the spin (single qubit) in a targeted quantum dot, we 
will modulate the Rashba component (and hence the total spin splitting energy)
 in that dot with a gate potential 
(applied through lithographically defined gates) to 
make it resonant with a {\it globally} applied ac magnetic field. This will 
induce 
transitions between the two spin states in that particular dot. Other dots 
remain unaffected. Depending on the amplitude 
of the magnetic field and duration of the interaction, the spin will rotate 
through chosen angles. 
The angle through which the spin rotates in a time $T$ is given by
\begin{equation}
\theta = {{1}\over{\hbar}} \int_0^T {\vec \mu_B} \cdot {\vec B} dt
\end{equation}
We will apply a potential 
pulse of width $T$ to the gate and this will rotate the spin through an angle 
$\theta$. For instance, a $\pi$ 
pulse ($\theta = \pi$) will 
flip the spin completely, whereas a $\pi$/2 pulse will place the
quanton in an equal superposition of the two spin states if it was originally in 
one of the spin eigenstates. Thus, arbitrary qubit
rotations in target dots can be achieved by controlling the 
width of the potential pulses applied at the gate and this realizes the 
first ingredient of a universal quantum gate, namely
arbitrary single qubit rotations.

\subsection{Two qubit entanglement:
Controlled dynamics of 2-qubit rotations}

In order to achieve the second and last ingredient of a universal 
quantum gate -- namely the conditional dynamics of a controlled NOT
operation -- we need to entangle 2 qubits and couple 
the rotation of one  (target qubit) with the orientation of the other (control 
qubit). This can be done by turning on an exchange coupling between 
two single quantons in two neighboring dots for specificed durations
of time. If the coupling is turned on a for a duration $h/2J$ ($J$ is
the exchange interaction energy), then the associated unitary time 
evolution corresponds to the ``swap'' operator which simply 
exchanges the quantum states of qubits 1 and 2. If the exchange is
turned on for one-half of that duration, then the unitary evolution
corresponds to a ``square-root of swap''. By applying a sequence of 
unitary operations corresponding to single qubit rotations and 
the square-root of swap operation, one can realize the controlled NOT
dynamics of a universal quantum gate \cite{loss}.

The exchange coupling between the control ($C$) and target ($T$)
dots can be varied by  
applying a positive potential (for electron qubits) or a 
negative potential (for hole qubits) to the contact $A_1$
which causes the electron- or hole-wave functions in $C$ and $T$ to 
leak out into the intervening barrier.
This  will turn the exchange coupling on. Thus, by applying appropriate
potential pulses to various gates in the set ($A_1$, $A_1$'), we can 
realize the complete universal quantum gate.

\subsection{Quantum connectors}

An essential ingredient in quantum computers is ``quantum connectors''
that transmit quantum information, much in the same way as normal metal wires
transmit voltages in a classical chip. A wire that transmits spin polarization
is simply a linear array of exchange coupled single-electron  quantum 
dots. The ground state of this array is anti-ferromagnetic so that the 
spin repeats itself in every other cell. Such as ``spin wire'' is shown in
Fig. \ref{schematic}c.

\subsection{Reading and writing of a qubit: spin polarizers and analyzers}

``Writing'' a qubit  involves assigning a predetermined polarization to the 
spin. Thus, we need a spin polarizer for writing. We can inject the 
quanton with a fixed spin from the polarizer, and then rotate it
by an appropriate angle to reach the desired initial orientation.
The polarizer can be a spin-polarized ferromagnetic contact, or even 
an ordinary metal contact if the energy states in the semiconductor 
dot are already spin split (see next section).

Reading a qubit involves measuring the spin orientation of the
qubit. This is 
achieved in the same way as in a spin transistor \cite{das}.
A suitable potential is applied between the ferromagnetic contacts to 
make the quanton trapped in the semiconductor dot flow out into 
a ferromagnetic contact that acts as a spin analyzer.
If the quanton's spin is oriented along the analyzer's magnetization,
the analyzer transmits the quanton; else it reflects it. A transmitted quanton
is therefore read as ``spin-up'', and the absence of a transmitted 
quanton is read as ``spin-down''. A qubit that has been 
read collapses to a classical binary bit with two possible 
values: ``spin-up'' and ``spin-down''. Single 
electron (or hole) currents or charge (due to a transmitted 
electron) can be detected by single electron electrometers
that are exquisitely sensitive and can detect the charge of 10$^{-8}$
electrons \cite{devoret}.

There are of course other strategies for reading a qubit, as mentioned before. 
It would 
cost more energy to inject a second quanton into a quantum dot
if its spin is parallel to the one already occupying the dot (Pauli
Exclusion Principle). The Pauli blockade will add to the Coulomb blockade
in this case and hence, in principle, can be detected in tunneling
current measurements. Methods using nanosized FETs have also been 
suggested for measuring single quanton spins \cite{vrijen}, but is a lot
more difficult than the present method. The problem with the Pauli
blockade approach is that one must know the spin of the first quanton
with certainty, in order to determine the spin of the second quanton.
The spin of the first quanton can be fixed with a highly localized 
magnetic field that {\it must not affect} the spin of the second quanton.
In principle, this can be done by generating a local magnetic field 
in a  nanometer sized area that is magnetically shielded. This is 
extremely difficult. In contrast, our technique of reading spin using 
spin analyzers is much simpler.

\section{Spin Coherence}

We have proposed using the Rashba effect to select a target qubit for rotation.
It is widely believed that this effect, which is   
an electric-field-induced spin orbit coupling, will degrade spin coherence 
 and be detrimental to quantum computing. That is true for bulk, quantum wells 
and quantum wires, but {\it not}
quantum dots. The use of quantum dots provides widespread immunity against
spin decoherence.

There are three principal mechanisms of spin relaxation in a semiconductor:
(i) the Elliott-Yafet mechanism, (ii) the D'yakonov-Perel mechanism,
and (iii) the Bir-Aronov-Pikus mechanism \cite{dassarma}. 

The Elliott-Yafet
mechanism \cite{elliott} is based on the fact that in a real crystal, the 
Bloch states are not spin eigenstates. The spin orbit interaction mixes
the spin-up and spin-down amplitudes. The Rashba effect {\it can}
exacerbate this mixing. As a result of this mixing, ordinary scattering 
interactions
with impurities, phonons, surfaces, etc., can flip spin. In a quantum dot
however, the electron is completely stationary since the wavefunction is
fully {\it real} and hence the expected value of the momentum in any 
direction is exactly zero. Since the electron does not move, it cannot
suffer elastic collisions which changes the electron's momentum
(since it cannot change energy). The only problem is therefore 
inelastic scattering (phonons). The Overhauser effect \cite{overhauser}
is known to flip spin via phonon modulated scattering. Thus, phonons could be
a nuisance. 

In the context of quantum computing however, phonons are almost never a serious 
issue since
the ambient temperature is usually very low (typically 300 mK - 4.2 K).
Moreover, in a quantum dot, the phonon bottleneck effect mitigates 
phonon scattering to a large extent \cite{benisty}. Thus, the Elliott-Yafet
mechanism is very weak. A far more important decohering mechanism is 
electromagnetic
interference from the gate. This is unavoidable in any gate controlled 
scheme. Again, the quantum dot helps in countering electromagnetic decoherence
since it usually needs a phonon or infrared photon to carry away the dissipative
energy. Because of the extremely restrictive selection rules for photon 
or phonon transitions in a quantum dot, the electromagnetic decoherence 
may be suppressed as well.

Another common spin decoherence mechanism (which dominates in semiconductor 
quantum wells or quantum wires where multiple subbands are occupied) is the 
D'yakonov-Perel mechanism \cite{dyakonov}. An electron's spin precesses about 
the direction of its momentum, as long as the crystal lacks inversion symmetry 
and/or has a structural inversion asymmetry (such as due to a built-in electric 
field caused by doping gradients, compositional modulation, or an 
externally applied electric field). The crystal inversion asymmetry gives rise 
to Dresselhaus spin orbit interaction \cite{dresselhaus} and the structural 
inversion asymmetry gives rise to Rashba spin orbit interaction \cite{rashba}. 
These spin orbit interactions are momentum dependent. They cause the spin to 
precess about an axis determined by the direction of the momentum. If the 
momentum changes randomly because of scattering, then the spin orientation 
changes randomly. When  ensemble averaged over a large number of 
electrons, the ensemble average spin dephases \cite{pramanik}. This 
dephasing mechanism is the so-called D'yakonov-Perel' mechanism.
 In a quantum dot, there is no ``momentum'', and there is no 
 ``ensemble'' of electrons either (we are talking of a single electron
 encoding the qubit). Therefore, the Dyakonov-Perel 
 mechanism does not exist. 
 
 What could be the most serious source of decoherence in a quantum dot
 is interactions with nuclear spins \cite{khaetski}. Therefore, one 
 should choose the right material for the quantum dot; it should be isotopically 
pure and the nuclear spin will be small. 
 
 The last mechanism of concern is the Bir-Aronov-Pikus mechanism \cite{bir} 
 which is based on electron-hole interaction. Since we are dealing 
 with a single conduction band electron, injected from the contact
 into a depleted semiconductor, this mechanism is extremely weak.
 
 In conclusion, the quantum dot is very forgiving in terms of spin coherence.
 
 In the contex of a quantum ``computer'' where several qubits are 
 entangled (and therefore interact), there may be more subtle decoherence 
 issues associated with many body effects. Dealing with decoherence 
 in the interacting electron picture is rather challenging, and so far 
 only few theoretical attempts have been made in that direction 
\cite{fedichkin}.

\section{Coherent spin injection from spin polarized contacts}

We mentioned earlier that we wish to inject an electron into the 
quantum dot with a definite spin orientation so that we know
the initial state of the qubit at time $t$ =0.  However,
coherent spin injection from a spin-polarized 
(ferromagnetic) contact into the semiconductor is {\it not} critical for this
purpose.
If the spin-injection does not work, one could still select a definite initial 
spin using the principle of the spin-RTD.
Since the subbands in the quantum dot are spin split, one can align the 
Fermi level in the contact with one of the spin-split levels and therefore
preferentially inject into that level. This will assign a definite 
spin orientation to the qubit. If even that does not work,
one can inject  an electron with arbitrary spin and wait till the electron 
decays 
to the lowest state (by a spin flip transition if necessary). Since 
the spin degeneracy is lifted by a combination of the Zeeman and 
Rashba effect (see Appendix), the ground state always 
has a definite spin polarization. In the latter two scenarios, ferromagnetic 
contacts are not necessary.

Even though coherent spin ``injection'' is not critical, coherent spin 
``detection'' is
absolutely critical since it provides the mechanism for reading a qubit. The 
injector
is a spin-polarizer and the detector is a spin-analyzer. We can do away
with the polarizer, but not the analyzer. Thus, coherent spin injection across
one of the ferromagnetic interfaces is necessary. If the analyzer does 
not work either, we have to rely on the mechanisms proposed in ref. 
\cite{vrijen} for reading spin which are much more complicated.

Coherent spin injection from a metal into a semiconductor is a difficult problem
\cite{tang} and has been recently addressed theoretically \cite{schmidt, 
rashba1}. It has been realized that efficient spin injection from a 
metallic ferromagnet into a semiconductor requires interposing a tunnel 
barrier of some sort (e.g. a Schottky barrier) between the two.
There have been some scattered reports of spin injection from a metallic 
ferromagnet
into a semiconductor \cite{bennett, hanicki} and between two semiconductors 
of widely different bandgaps \cite{awschalom}.
Recently, spin polarized hole injection was demonstrated from
GaMnAs into GaAs \cite{ohno} at around a temperature of 120 K. Prior to that, 
spin polarized injection from CdMnTe into CdTe was demonstrated
\cite{oestreich}, but the disadvantage in that case is that CdMnTe is not
a permanent ferromagnet; the
spin polarization needs to be maintained by  a globally applied 
dc magnetic field which introduces a Zeeman splitting in CdMnTe. However,
only a very small field is required since the effective Land\'e g-factor
for electrons in dilute magnetic semiconductors is huge ($\sim$ 100).
On the other hand, the advantage of CdMnTe is that it
is lattice matched to CdTe and hence interface scattering is 
less of a problem.  Most recently,  90\% spin polarized electron injection 
was demonstrated from the dilute magnetic semiconductor 
Be$_x$Mn$_y$Zn$_{1-x-y}$Se into GaAs 
at a temperature below 5 K \cite{fiederling} and at a relatively
large magnetic field which induces a Zeeman splitting in the magnetic
semiconductor. While the temperature is high enough for quantum
computing applications, the applied magnetic field is too large
and may flip the spin in the semiconductor quantum dot, thus 
corrupting the qubit. The problem of coherent spin injection 
from a ferromagnetic material into a semiconductor is a topic
of much current research. It has a long history and rapid strides are
being made in this field.

Another important question is how easy will it be to maintain 
single electron occupancy in each dot. As long as the energy cost to add an 
additional electron (= $e^2/2C$; $C$ is the capacitance of the dot)
significantly exceeds the thermal energy kT, only a single electron
will occupy each dot. Uniform 
electron occupancy in arrays of $>$ 10$^{8}$ dots has been 
shown experimentally \cite{muerer}.

\subsection{Spin measurement}

After quantum computation is over, we need to read the result by measuring
the qubits. During this process, the qubits will collapse to classical 
bits. These classical bits are the measured spin orientations in relevant
dots. They are measured    
by measuring the current that results when the potential over the 
dot is raised over the Coulomb blockade threshold. If we assume 
that the differential phase-shift suffered by the spin in traversing the
dot is negligible; in other words, transport through the dot does not rotate 
the spin, then the magnitude of the measured current can tell us the
spin orientation \cite{das}. It was shown in ref. \cite{das} that 
the spin-polarized contacts act as electronic analogs
of optical polarizers and analyzers, so the current will depend on the 
projection of
the spin of the quantum dot's resident electron on the spin orientation in the  
contacts. Thus, by 
measuring the current, we can tell the spin orientation in any quantum 
dot. 

\subsection{Calibration}

For each target dot, the gate
potential $V_{target}$ that needs to be applied to bring the spin-splitting 
energy in resonance with $B_{ac}$, can be calibrated following the 
procedure outlined by Kane \cite{kane}. With $B_{ac}$ = 0, we measure 
the spin in a quantum  
dot. Then we switch on $B_{ac}$ and sweep $V$ over a range. Next $B_{ac}$
is switched off and the spin is measured. The range of $V$ is progressively 
increased
till we find that the spin has flipped. We then proceed to narrow the 
range with successive iteration while making sure that the spin
does flip in each iteration. Finally this allows us to ascertain $V$
with an arbitrary degree of accuracy. As pointed out by Kane \cite{kane}, 
the calibration procedure can, in principle, be carried out in parallel
over several dots simultaneously and the voltages stored in adjacent
capacitors. External circuitry will thus be needed only to control 
the timing of the biases (application of $V_{target}$) and not their
magnitudes. While this is definitely an advantage, fabricating nanoscale 
capacitors adjacent to each individual 
dot is certainly not easy. Moreover, capacitors 
discharge over time, requiring frequent
recalibration through refresh cycles;
therefore, this may not be a significant advantage.

\subsection{Input and Output Operations}

Any computer is of course useless unless we are able to input and output
data successfully. Since we are using spin-polarized contacts to 
inject an electron in each dot, we know the initial orientation. Those
dots where the initial orientation is the one we want are left unperturbed
while the spins in the remaining dots are flipped by resonating with $B_{ac}$. 
This process prepares
the quantum computer in the initial state for a computation and can be 
viewed as the act of ``writing'' the input data.
Computation
then proceeds on this initial state by carrying out a desired sequence
of controlled rotation operations. Reading the data is achieved as described
before.

\section{Experimental realization of a quantum gate}

\begin{figure}[h]
\epsfxsize=4.3in
\epsfysize=2.9in
\centerline{\epsffile{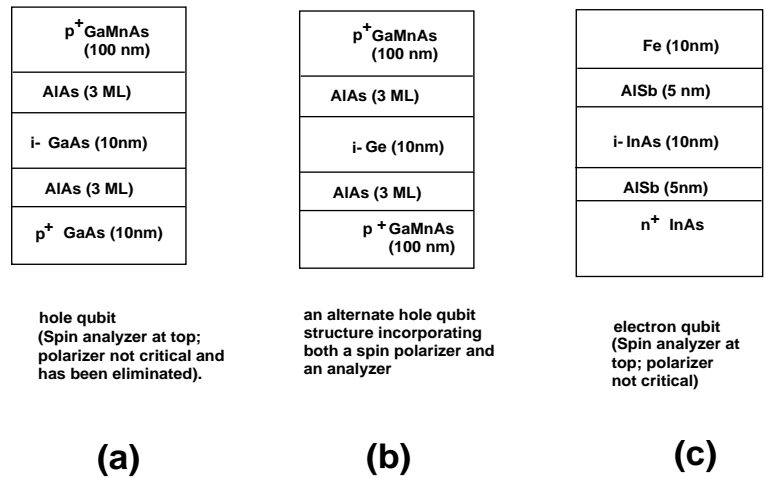}}
\caption[]{\tt MBE-grown multilayered films for qubits: (a) a ``hole-qubit''
structure in which the GaMnAs ferromagnetic layer in the spin analyzer. The host 
for the qubit (a single hole) is the GaAs layer. This structure 
is the easiest to grow and has been demonstrated by Jonker, (b) 
an improved ``hole-qubit'' structure consisting of a GaMnAs polarizer and 
analyzer layer. The host of the qubit is Ge which has a smaller hole effective 
mass than GaAs and hence a stronger Rashba coupling for holes, (c) 
an ``electron qubit'' structure with Fe as the spin analyzer layer.
\label{MBE}}
\end{figure}

A team of researchers from the author's group and the US Naval Research 
Laboratory have proposed three different structures grown by molecular 
beam epitaxy that are viable prototypes for a quantum gate. They are shown in 
Fig. \ref{MBE}.

The first two structures are for hole 
qubits and the last for an electron qubit. Each structure has at least one 
ferromagnetic layer that acts as a spin analyzer. A variant of the first 
structure was used 
in the past to demonstrate spin coherent 
injection of {\it holes} from GaMnAs (a ferromagnet below 120 K) into GaAs 
\cite{ohno}. In these structures, GaMnAs is used as a ``spin analyzer'',
rather than a ``spin polarizer'', since the spin analyzer is more critical. 

The hole qubit structure shown in Fig. \ref{MBE}(b) uses Ge as the 
semiconductor
host for the  qubit because Ge has a smaller hole effective mass 
than GaAs. The Rashba spin splitting energy is inversely proportional to the 
effective mass of the quanton  which motivates the choice
of Ge. The use of Ge as opposed to GaAs has other advantages. The larger
valence band offsets result in stronger hole confinement. Moreover,
the hole spin coherence time in Ge, an elemental semiconductor,
is probably {\it much longer} than in GaAs.
Ge is essentially
lattice matched to GaAs and high quality epilayers have been grown.
Since Ge can be grown at low temperatures (compatible with GaMnAs),
this structure can have two ferromagnetic 
layers: one at the top and one at the bottom. {\it Thus, it has both 
a spin polarizer and an analyzer.}

A structure for ``electron-qubit'' is shown in Fig. \ref{MBE}(c). 
It consists of an InAs quantum well to act as the host for the qubit,
thin AlSb barriers, topped by a layer of Fe which serves as the 
spin analyzer. InAs is chosen because it has a large Land\'e g-factor
which causes a larger Zeeman splitting (and hence indirectly a larger
Rashba splitting (see Appendix)). It 
also has a small electron 
effective mass which directly enhances the Rashba splitting. It is the 
ideal material to host an electron qubit.

\subsection{Creation of buried quantum dots using shallow etching
and surface depletion}

\begin{figure}
\epsfxsize=3.4in
\epsfysize=3.4in
\centerline{\epsffile{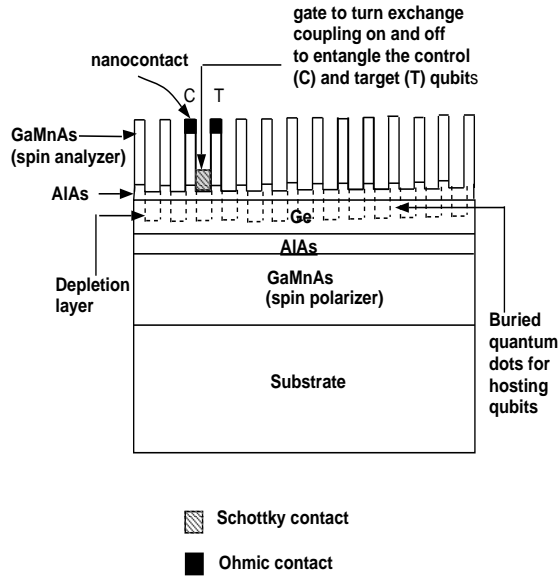}}
\caption[]{\tt Shallow etched pillars defined on MBE grown films
for creating buried quantum dots that host qubits. Reproduced 
with permission of Vasov Publishers from ref. \cite{plds}.
\label{etch}}
\end{figure}

The semiconductor quantum dots must be {\it buried} dots as opposed
to exposed dots for two reasons: (i) first, the interfaces of
buried dots (which are defined by electrostatic depletion) are
far superior to those of exposed dots, (ii) second, and more 
importantly, we need the wave functions of two neighboring 
dots to leak out into the intervening barrier and exchange-couple
under the influence of a gate potential. This is needed
in order to implement the ``square-root of swap'' operation
which is necessary for the universal controlled-NOT gate.
The barriers must be somewhat transparent for significant 
exchange coupling. These requirements are more easily met 
with ``buried'' dots.
Buried dots can be created by shallow etching (sputtering) using etch
masks that are delineated by either x-ray
lithography (using resists) or by self-assembly (resistless). A side 
view of the finished quantum gate is shown in Fig. \ref{etch}. 

\section{Quantum cryptography}

We now venture into a very different field, quite distinct from quantum 
computing, namely ``quantum cryptography''. Cryptography is the science of 
encoding confidential data in cryptic 
messages and sending it over public channels to an intended receiver.
An eavesdropper,  who may intercept the message, should not be able to decipher 
it.  Deciphering is accomplished via a ``key'' provided by the 
sender to the receiver alone. Nobody else should have this private key. In 
conventional classical cryptography, a ``public key'' is 
transmitted over a public channel and the actual ``private key'' 
is extracted from the public key by solving a trapdoor function such
as factorizing a large number into its prime constituents. Although,
solving the trapdoor function is difficult, it is theoretically {\it not 
impossible}. Worse still, if the eavesdropper succeeds in extracting 
the private key, neither the sender, nor the receiver will be aware 
of this. From then on, all secret transmissions will be compromised.

In the quantum world, there is no way to prevent an eavesdropper 
from intercepting the message, but once the eavesdropper does so,
both the sender and receiver will become aware of her presence.
Therefore, the intercepted key will be discarded and a new key 
will be sent out. The process will continue until an unintercepted 
key arrives at the receiver. The difference between quantum and 
classical cryptography is that in quantum cryptography, the 
eavesdropper can never remain hidden. Thus quantum cryptography 
is invariably more secure and unlike in the classical case, the 
final key is {\it guaranteed} to be secure.

Charles Bennett and Gilles Brassard devised a quantum 
cryptography scheme based on the quantum no-cloning theorem and 
the impossibility of measuring certain pairs of observables simultaneously 
\cite{bennet_crypt1, bennet_crypt2}. The latter is the Heisenberg uncertainty 
principle
which states that pairs of observables whose operators do not commute, cannot 
be measured simultaneously. The best known example of such pairs 
is the position and momentum of a quantum object (e.g. an electron).
We will call such pairs ``conjugate variables''.
The no-cloning theorem is an extension of quantum measurement theory.
In order to clone a quantum object, the cloning 
devise must first measure it or examine it. In measuring one attribute
of a quantum object (an observable), its conjugate pair partner is 
invariably disturbed, and disturbed unpredictably. Therefore, there is
no way to capture a quantum object with complete fidelity which is a 
pre-requisite for cloning.

\subsection{A quantum cryptosystem employing photons}

Consider a photon which can be polarized in the so-called rectilinear 
basis, or the diagonal basis. The rectilinear basis 
corresponds to the photon's electron field being aligned at 0$^{\circ}$
or 90$^{\circ}$ to some reference line on a plane, while the diagonal
basis corresponds the electric field being aligned at 45$^{\circ}$ or 
135$^{\circ}$ to the 
line.   The rectilinear
and diagonal polarizations are conjugate pairs and hence it is impossible 
to measure both of them exactly, at the same time.

In the humane convention of communication and information sciences, the 
sender A is always named Alice and the receiver B is always named Bob.
Alice encodes the bits that she wants to send to Bob in a photon's polarization 
by choosing
 either the rectilinear basis or the diagonal basis {\it randomly}. Her 
resulting polarization 
angle for each bit depends on the bit value and the chosen basis as follows: 

\bigskip

\begin{tabular}{|c|c|c|}
\hline
Bit value & Chosen basis & Polarization angle \\
\hline
0 & rectilinear & 0$^{\circ}$ \\
1 & rectilinear & 90$^{\circ}$ \\
0 & diagonal & 45$^{\circ}$ \\
1 & diagonal & 135$^{\circ}$ \\
\hline
\end{tabular}

\bigskip

When Bob receives the bits, he measures them also in either the rectilinear
or the diagonal basis randomly. For those bits where he and Alice coincidentally 
used the 
same bases, there should be perfect correlation. If Alice sent out a 
bit 1 in rectilinear basis, meaning a polarization angle of 90$^{\circ}$ (see 
above table), and Bob also happens to measure  it in the rectilinear basis, then 
he should 
measure an angle of 90$^{\circ}$ and interpret the bit as 1. 

Alice and Bob then announce over a public (insecure) channel what bases 
they used for each bit measurement. They then discard those bits where they
happened to have used {\it different} bases. In the case of the remaining
bits, where they accidentally used the {\it same} bases, Alice can predict 
perfectly
what bit value Bob should have measured each time. She sent those bits out and 
Bob's measurement scheme was correct by happenstance; so there should be perfect 
agreement. Similarly Bob can 
predict what bits Alice sent out, because he measured them and he knows that he 
measured
them correctly. The only exception to this happens if an eavesdropper
(named Eve) is lurking in the channel and disturbed the polarizations 
in her attempt to eavesdrop on the bits Alice was sending out.
We will show in the next section that quantum mechanics (or more precisely 
the Heisenberg uncertainty principle and the no cloning theorem) guarantees that 
this
disturbance must take place whenever Eve intercepts a photon.

Alice now announces over a public channel {\it some} of the bits she sent out
and asks Bob to check if he received those particular bits intact. If he did, 
then there 
is likely to be no eavesdropper and the channel is secure. Any disagreement
between the values of the bits Alice sent out and Bob measured correctly (with 
the same basis) reveals the presence of an eavesdropper who caused a 
disturbance. In the quantum world, Eve cannot hide since she cannot avoid
causing a disturbance whenever she tries to eavesdrop on the bits.

Here is why Eve causes a disturbance. In her attempt to eavesdrop, she 
must intercept the bits from Alice before they reach Bob and measure 
the polarizations to decipher the bits. Since she does not know what
bases Alice used to encode the bits, she measures them in random bases.
Let us consider the situation when Alice sent out a bit with value 1
in the rectilinear basis and Eve measured it in the diagonal basis. Because
of the uncertainty relationship between the rectilinear and diagonal 
bases, Eve cannot tell whether the bit was 0 or 1.  In her attempt to hide,
she must  send a photon out to Bob pretending to be Alice. One half of the 
time she will send out the incorrect bit 0 in the rectilinear basis (this 
is also the no-cloning theorem; she can never clone the photon's polarization -  
a quantum entity - with 100\% fidelity).
Let us say now that Bob measures this in the rectilinear basis and measures a 
0 because that is what he received from Eve. Later, when Bob and Alice
compare notes over this bit, they will realize that Alice sent a bit with value 
1  but 
Bob received a 0. Therefore, this disagreement reveals the presence of 
Eve.

Of course, just a few disagreements do not reveal the presence of Eve
since they could be due to imperfections in the channel or other 
errors that cause bit flips. However, consistent disagreements in 
several bits reveal the presence of Eve. In fact, the probability of detecting
Eve goes as 1 - (3/4)$^n$ where $n$ is the number of bits
tested. Thus, Eve can be detected with near unit probability by testing 
a large number of bits. Once Eve is detected, Alice and Bob stop communicating
and try again later. This goes on till Eve gets bored and leaves.

In this section, we showed that quantum mechanics guarantees secure 
cryptography. In 1989, IBM built a quantum cryptography machine 
\cite{bennet_crypt3}. Today,
there is a 30 km long quantum cryptography line operating under Lake Geneva
in Switzerland \cite{marand}.

\section{Quantum teleportation}

Anybody familiar with the popular sci-fi series Star Trek knowns 
what teleportation is supposed to do. It takes an object (a human
or a klingon in Star Trek), destroys it at the source, and 
recreates it {\it exactly} at the destination. The object does
not traverse the intervening distance and hence the speed of 
teleportation, in principle, need not be restricted by the speed
of light.

In the classical world of today, if we want to transmit an object
such as this page to a reader some distance away, we will probably
use a facsimile machine. This is very different from teleportation
in two ways: first, the original is not destroyed at the source,
and second, what is received at the destination cannot be an {\it exact}
replica. It cannot be exact because of quantum mechanics. No matter 
how sophisticated the facsimile machine is, the individual lettering
in this page is made up of atoms and sub-atomic particles. When the 
light from the facsimile machine shines on them to determine their
exact position on the page, they impart them with a random momentum
which will displace them randomly. 
This is Heisenberg's uncertainty
principle at work which states that it is impossible to specify
simultaneously two quantities whose operators do not commute, such as position 
and momentum, precisely. Thus, the lettering that is transmitted
is never exactly what was fed to the facsimile machine. There must be
some distortion.

What is true of a classical object such as this page of text,
is even more true of a quantum object. Generally speaking,
 it is impossible to
measure all the attributes of a quantum state simultaneously and
exactly. Therefore, it is impossible to 
clone a quantum object. This is the basis of the {\it quantum no-cloning 
theorem}. 

Even though quantum mechanics forbids cloning or copying, it does not
forbid teleportation where the original object is sent {\it absolutely intact}
to the destination, but is then unavailable (effectively destroyed) at the 
source. Achieving this feat requires the use of Einstein-Podolsky-Rosen
pairs which are twin quantum objects that are forever correlated in some way.
This correlation or ``entanglement'' can occur if the two quantum 
objects have interacted at some time in the past.

Albert Einstein, Boris Podolsky and Nathan Rosen (henceforth 
called EPR) came up with the 
following gedanken experiment to question the validity of quantum
mechanics \cite{einstein}. Imagine the two electrons of a hydrogen
molecule. These two electrons have interacted to form the molecule
and are in the bonding state where their spins are known to be 
anti-parallel (singlet state). Now, an explosion rips the molecule
apart and sends one electron to New York and the other to Tokyo.
At this time, we know that the two electrons have opposite 
spins, but we do not know which one has upspin (measured in some basis)
and which one has downspin. We will now attach 
labels to the two electrons (electrons are still treated as 
equals and indistinguishable particles) and denote the Tokyo electron
with subscript 1 and the New York electron with subscript 2.
The ``spin part'' of the actual two electron wave function
must then be written as 
\begin{equation}
\psi = {{1}\over{\sqrt{2}}} (| \uparrow_1 \downarrow_2 > \pm |\downarrow_1 
\uparrow_2 >)
\label{entangled}
\end{equation}
This wave function indicates that either the Tokyo electron's spin is ``up'' and 
the New York electron's spin is ``down'', or vice versa. These two possibilities 
have
exactly the same probability. Note that even though we do not
know the spin of either electron, we know that they are always 
anti-parallel. Therefore, the spins of these two electrons, separated 
by the vast Pacific Ocean,  
have become entangled like Siamese twins.

EPR was interested in a paradox that questioned the completeness of quantum
theory. Let us concentrate on the
electron in Tokyo and forget about the one in New York for a moment.
The spin part of the Tokyo electron's wave function is aptly 
written as 
\begin{equation}
\psi_{Tokyo} = {{1}\over{\sqrt{2}}} ( |\uparrow_1 > \pm |\downarrow_1> )
\label{tokyo}
\end{equation}
That is, its spin has equal probability of being ``up'' and ``down''.
Now, let us say, we measure the spin in New York using a Stern Gerlach 
apparatus and find it to be ``up''. We will immediately know that the 
Tokyo electron's spin is ``down''. We have therefore collapsed the Tokyo
electron's spin wave function (given by equation \ref{tokyo}) to the ``down'' 
state, and at the same time, 
received information about the spin (namely , that it is ``down''), by 
performing a
measurement in New York. The information about the Tokyo electron's
spin has therefore traveled to New York instantaneously with 
infinite speed thereby breaking the light speed barrier. EPR
argued that this is impossible and therefore quantum mechanics
must be incomplete.

To remedy the situation, it was proposed that there must be local
hidden variables in quantum mechanics which, when properly
accounted for, can solve this conundrum. But in 1964, Bell proved 
a mathematical inequality, that showed that there are no {\it local} 
hidden variables in quantum mechanics \cite{bell}. However, there
can be {\it non-local} hidden variables that could allow violation 
of Bell's inequality and reconcile the
EPR paradox with quantum mechanics. In 1982, Aspect, Dalibard and Roger,
demonstrated the violation of Bell's inequality using entangled 
photon pairs (whose polarizations are entangled), thereby demonstrating 
non-local hidden variables. Collapse of the wave function is a very
unique type of interaction that is non-local unlike the known forces
in nature (electroweak, gravity, and strong interactions).
The predictions of quantum mechanics 
have never been proven false. 

\subsection{Entangled states}

Mathematically, the wave functions of the entangled state of two particles
are such that they cannot be written as the product of the wave functions
of each particle. Recall the EPR spin pairs in New York and Tokyo. The wave 
function of the Tokyo electron's spin is given by Equation \ref{tokyo}
and similarly the wave function of the New York electron's spin
will be given by

\begin{equation}
\psi_{Tokyo} = {{1}\over{\sqrt{2}}} ( |\uparrow_2 > \pm |\downarrow_2> )
\label{ny}
\end{equation} 

Note that the product of Equation \ref{tokyo} and \ref{ny} does not yield 
Equation \ref{entangled}. The wave function of the entangled state of 
two objects cannot be written as the product of the wave functions of
each one of them. 

\subsection{Teleporting a single qubit}

In 1993, Charles Bennet and his co-workers showed how entangled 
EPR pairs can be used to teleport a quantum object or more correctly,
the quantum attribute of an object, over arbitrary distances 
\cite{bennet_teleport}. This 
will not allow us to teleport an entire electron physically, but we can
teleport its {\it spin state} to another electron at a remote location.
Since electrons are indistinguishable particles, this is indeed tantamount to 
teleporting the electron.

Consider the case where a qubit is encoded in the spin of a single
electron (particle 1). The spin wave function or qubit is written as 
\begin{equation}
\phi_1 = a |\uparrow_1> + b |\downarrow_1> = \left ( \begin{array}{c}
a \\
b \\
\end{array} \right )
\label{one}
\end{equation}
where $|a|^2 + |b|^2$ = 1.

To teleport this object from a sender Alice to a receiver Bob, we will
need Alice and Bob to have shared in the past an EPR pair which we will call 
particles
2 and 3. Particle 2 is with Alice and particle 3 resides with Bob.
Their joint state wave function 
can be written as Equation \ref{entangled}
with just the labels altered.
\begin{equation}
\phi_{23} = {{1}\over{\sqrt{2}}} (| \uparrow_2  \downarrow_3 > - |\downarrow_2 
\uparrow_3 >)
\end{equation}

At this point, particle 1 is still uncorrelated with particles 2 and 3.
Therefore, we can write the three-particle (or three-spin) wave function as a 
product between $\phi_1$ and $\phi_{23}$
\begin{eqnarray}
\phi_{123} & = &  \phi_1 \otimes \phi_{23} \nonumber \\
& = & \phi_1 \otimes {{1}\over{\sqrt{2}}} (| \uparrow_2 \downarrow_3 > - 
|\downarrow_2 \uparrow_3 >) \nonumber \\
& = & {{a}\over{\sqrt{2}}} (|\uparrow_1> \otimes |\uparrow_2> \otimes 
|\downarrow_3> - |\uparrow_1> \otimes |\downarrow_2> \otimes |\uparrow_3> 
\nonumber \\
& + & {{b}\over{\sqrt{2}}} (|\downarrow_1> \otimes |\uparrow_2> \otimes 
|\downarrow_3> - |\downarrow_1> \otimes |\downarrow_2> \otimes |\uparrow_3>
\end{eqnarray}

Alice now makes a so-called Bell measurement to entangle the qubit 
(particle 1) with her particle 2. She lowers the potential 
barriers separating particles 1 and 2 and makes their 
wave functions overlap in space. The subsequent Bell measurement will involve 
measuring the total spin of particles 1 and 2, and the symmetry of the
spatial parts of their coupled  wave functions, that is whether the spatial 
part 
is symmetric in space or anti-symmetric. Typically, the symmetric
eigenstate will have lower energy than the anti-symmetric eigenstate
so that a measurement of the energy eigenstate is effectively a measurement 
of the spatial symmetry.

The wave function of the 
joint states of particles 1 and 2 after they are entangled
(i.e. after their wave functions are made to overlap) can have one of
only four forms. They can be written as 
\begin{eqnarray}
\psi_A = {{1}\over{\sqrt{2}}} (| \uparrow_1 \downarrow_2 > - |\downarrow_1 
\uparrow_2 >) \nonumber \\
\psi_B = {{1}\over{\sqrt{2}}} (| \uparrow_1 \downarrow_2 > + |\downarrow_1 
\uparrow_2 >) \nonumber \\
\psi_C = {{1}\over{\sqrt{2}}} (| \uparrow_1 \uparrow_2 > - |\downarrow_1 
\downarrow_2 >) \nonumber \\
\psi_C = {{1}\over{\sqrt{2}}} (| \uparrow_1 \uparrow_2 > + |\downarrow_1 
\downarrow_2 >)
\label{bell}
\end{eqnarray}

These states are called Bell basis states and they are mutually orthogonal.
Note that the first two states are singlet and the last two are triplet.
Hence a measurement of total spin will separate these two subgroups.
Then within each subgroup, one state is symmetric in spin and the other 
is antisymmetric. Since the total two-body wave function (spin + spatial part)
must be anti-symmetric, a symmetric state in spin implies that the 
spatial part of the wave function must be anti-symmetric and vice versa.
Thus measurement of the symmetry of the spatial part and the total 
spin is sufficient to identify the actual state among the four given 
in Equation \ref{bell}. The purpose of Alice's Bell measurement is
to complete this unambiguous identification.

Using Equation \ref{bell}, we can re-write Equation 55 as 
\begin{eqnarray}
\phi_{123} & = &  {{1}\over{2}} [ \psi_A(-a |\uparrow_3> - b 
|\downarrow_3> ) +
\psi_B ( -a |\uparrow_3>  + b |\downarrow_3> ) \nonumber \\
& + & \psi_C(a |\downarrow_3> + b |\uparrow_3> ) + \psi_D (a |\downarrow_3>
-b |\uparrow_3> ) ]
\end{eqnarray}

Using the column vector notation of Equation \ref{one}, we can 
re-write the above equation as
\begin{equation}
\phi_{123} = {{1}\over{2}} \left [ \psi_A \left ( \begin{array}{c}
-a \\
-b \\
\end{array} \right )_3
+ \psi_B \left ( \begin{array}{c}
-a \\
b \\
\end{array} \right )_3
+ \psi_C \left ( \begin{array}{c}
b \\
a \\
\end{array} \right )_3
+ \psi_D \left ( \begin{array}{c}
-b \\
a \\
\end{array} \right )_3 \right ]
\end{equation}

The 3-particle state is now such that if Alice's Bell measurement reveals
that particles 1 and 2 are in state $\psi_A$, then Bob's particle 
3 encodes the qubit $ (-a |\uparrow> - b \downarrow>)$. if Alice's measurement 
reveals the state $\psi_B$, then Bob's particle 
3 encodes the qubit $ (-a |\uparrow> + b \downarrow>)$, if Alice's measurement 
revelas the state $\psi_C$, then Bob's particle 
3 encodes the qubit $( b |\uparrow> + a \downarrow>)$, and finally if
 Alice's measurement 
revelas the state $\psi_D$, then Bob's particle 
3 encodes the qubit $(- b |\uparrow> + a \downarrow>)$.

After the Bell measurement, Alice will know how the state of Bob's particle
3 is related to the original qubit (particle 1), but Bob does not.
So Alice
must send Bob the result of the Bell measurement. Since there are 
four possible outcomes of the Bell measurement, Alice needs to send
Bob two classical bits telling him which outcome was observed. Accordingly,
Bob applies a unitary transformation to particle 3 to re-create particle 1
and hence the qubit. These unitary transformations are
\begin{eqnarray}
\psi_A & \rightarrow &  
\left ( \begin{array}{cc}
-1 & 0 \\
0 & -1 
\end{array}
\right ) \nonumber \\
\psi_B & \rightarrow &  
\left ( \begin{array}{cc}
-1 & 0 \\
0 & 1 
\end{array}
\right ) \nonumber \\
\psi_C & \rightarrow &  
\left ( \begin{array}{cc}
0 & 1 \\
1 & 0 
\end{array}
\right ) \nonumber \\
\psi_D & \rightarrow &  
\left ( \begin{array}{cc}
0 & -1 \\
1 & 0 
\end{array}
\right )
\end{eqnarray}

Note that since Alice must send Bob 2 classical bits to help him transform 
his particle 3 to particle 1, particle 1 was not teleported to Bob instantly.
There was no superluminal signaling. The classical bits must travel to
Bob with speed equal to or less than the speed of light in vacuum. There is 
one subtle exception however. One fourth of the time (corresponding to 
$\psi_A$), Bob does not have to apply any transformation (apart from an 
irrelevant phase factor of $\pi$) to his particle 3 in order to re-create
particle 1. Therefore, if Alice and Bob are content to live with a 
25\% success rate, instant teleportation is possible. It does not violate
relativistic dictates since no definite information is being transmitted.
In fact, the 25\% teleportation has been demonstrated experimentally by 
Zeilinger's
group in Austria.

\section{Quantum memory}

No review article is ever complete. But this review will be seriously amiss
if we do not mention a few words about quantum ``memory''. Most of the 
attention in the field of quantum computing has be focused on quantum logic, and 
very little 
on quantum memory.

Quantum memory is an important constituent of quantum information 
science. It has many applications: (i) increasing the efficiency
of quantum key distribution (QKD) protocols (the receiver Bob 
stores the received qubits in a quantum memory and measures 
them {\it after} the sender Alice tells him the bases), (ii) improving
the EPR-based QKD schemes \cite{ekert}, (iii) teleporting a state using 
singlet pairs prepared in advance, (iv) new schemes for QKD 
that rely on the existence of short-term memory \cite{bennet_mem, goldenberg},
(v) attacking oblivious transfer and quantum bit commitment schemes \cite{mor},
etc.

The requirements for quantum memory are thought to be 
very different from those of quantum 
gates.  In a quantum gate, the qubits are accessed 
and rotated numerous times, but the coherence time need 
not be very long; it simply has to be  
much longer than the switching time. In contrast, the qubits in a quantum 
memory are seldom accessed, but they must live much longer 
(ideally ``forever'') without decohering.
One must also be able to access them with high fidelity.

Since this review has focused on spintronic solid-state realizations of
quantum logic, we will stick to the spintronic solid state realization 
of quantum memory. This is no easy task since spin coherence times,
although longer than charge coherence times, are still very short. 
As a result, {\it non-volatile} quantum static memories (Q-SRAMs) are not
appropriate; rather, quantum
{\it dynamic memories} (Q-DRAMs) may be possible if the qubit can be {\it 
refreshed} 
periodically  through refresh cycles. Below, we explore possible 
routes to refreshing the quantum state of a quanton \cite{bandy_ecs}.

\subsection{Refreshing a qubit}

It is very possible that 
refreshing can be accomplished through the  {\it quantum
Zeno effect} which postulates that repeated observations 
 of a qubit will inhibit its decay \cite{sudarshan,
gurvitz}. Repeated observations automatically serve as  refresh cycles. However, 
this repeated observation has to be carried out
by a non-invasive detector. A ballistic point contact has been 
used in the past as a non-invasive charge detector for electrons 
in quantum dots \cite{field}, and its role in the
context of the quantum Zeno effect has been examined \cite{gurvitz}.
It may be possible to use a spin-polarized scanning tunneling
microscope tip as a non-invasive probe for spin, but that is yet to 
be realized in practice.

But what happens if the probe is an ``invasive'' probe? All is not lost. It may 
be possible to re-create
the entire 
qubit (including the phase) after it has ``interacted'' with the probe. This 
involves  {\it quantum
erasure} \cite{hillary, scully, kwiat} as explained below.

Consider a quanton in a coherent superposition of two spin states,
described by a wave function
\begin{equation}
\psi = a_{\uparrow} |\uparrow> + a_{\downarrow} |\downarrow>
\end{equation} 

A fundamental result of quantum measurement theory is that if
the spin analyzer tries to detect the spin of the incoming quanton,
the interaction between the detector and the quanton causes the wave function of 
the detector to become entangled with that 
of the quanton. The entangled (non-factorizable) wave function
is
\begin{equation}
\Phi = a_{\uparrow} |\uparrow> |1> + a_{\downarrow} |\downarrow> |2>
\end{equation}
where the wave functions $|1>$ and $|2>$ span the Hilbert space 
of the detector. Thus, $|1>$ corresponds to the detector (spin-analyzer)
passing an up-spin quanton, and $|2>$ corresponds to the
detector reflecting a downspin quanton.

If we make a measurement of whether the detector passed the quanton
(corresponding to the determination that the quanton's spin was ``up''),
the probability amplitude of that is
\begin{equation}
\Psi = <1|\Phi> = a_{\uparrow} |\uparrow> <1|1> + a_{\downarrow} |\downarrow> 
<1|2>
\end{equation}
Since the detector makes an  ``unambiguous'' determination, meaning that 
it {\it always} passes an upspin quanton and {\it never} passes a
downspin quanton, the wave functions $|1>$ and $|2>$ are orthogonal,
meaning that upspin detection and downspin detection are mutually
exclusive (a quanton cannot be simultaneously both upspin and downspin,
and the detector will unambiguously determine what the spin is).
Hence, from Equation (62),
\begin{equation}
\Psi_{detected} = a_{\uparrow} |\uparrow>; ~~~~~ |\Psi|^2 = |a_{\uparrow}|^2
\end{equation}
and we get no information about $a_{\downarrow}$, or the phase. 
This is interpreted as wave function collapse. However, the entanglement
of the detector with the quanton (Equation (61)) does not itself cause 
irreversible collapse. It
is not irreversible since if we design an experiment whose 
result is the probability of a particular outcome of the spin measurement
{\it and} finding the detector in the symmetric state ($|1> + |2>$),
then the corresponding probability amplitude is 
\begin{eqnarray}
[<1| + <2|]|\Phi> & = & a_{\uparrow} |\uparrow> <1|1> + a_{\downarrow} 
|\downarrow> <1|2> + a_{\downarrow} |\downarrow> <2|2> + a_{\uparrow} |\uparrow> 
<2|1> \nonumber \\
& = & a_{\uparrow} |\uparrow> + a_{\downarrow} |\downarrow> \nonumber \\
& = & \psi ~,
\end{eqnarray}
which is the original wave function. Hence, we can restore the original 
wave function $\psi$ from the entangled wavefunction $\Phi$. This is possible 
since Equation (61) still represents a ``pure'' state and not a ``mixed'' state.

Note that if we can find the detector in the symmetric state, we would 
not have known whether the quanton that passed through it was ``up'' 
or ``down'', and hence we would not have collapsed the wave function.
Thus, by finding the detector in the symmetric state, we have 
foregone any information about the spin and hence restored the 
original coherent superposition state from the entangled state (quanton
entangled with detector).
The quantum erasure is possible because the entangled wave function 
$\Phi$ is still a pure state and not a mixed state.

\begin{figure}
\epsfxsize=4.3in
\epsfysize=3.4in
\centerline{\epsffile{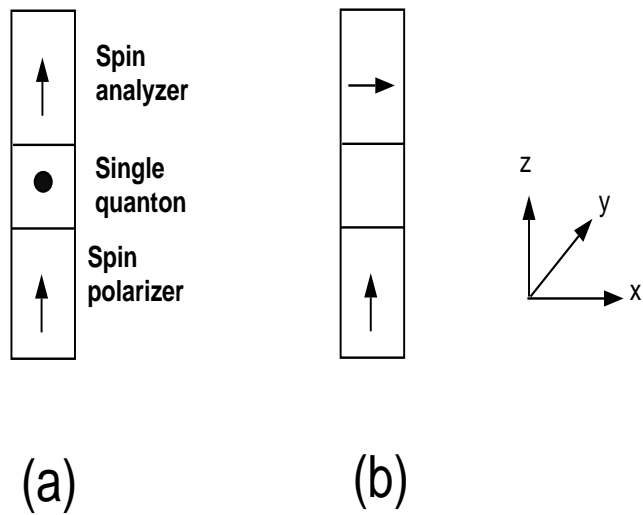}}
\caption[]{\tt The (a) initial and (b) final state of the polarizer-analyzer 
combination after the passage of a quanton corresponding to the 
reading of a qubit. Reproduced from ref. \cite{bandy_ecs} with permission of 
the Electrochemical Society. \label{erasure}}
\end{figure}

What do we need to implement quantum erasure? We need only one difficult 
and ambitious technological feat. When the 
quanton passes through the spin analyzer, it should be able to 
rotate the magnetization of the analyzer and {\it change it} in a particular 
way.
 If the
polarizer and analyzer were originally magnetized in the +z-direction, the 
passage of the quanton through the analyzer must turn on some 
interaction that results in the analyzer getting magnetized in the 
+x-direction. Fig. \ref{erasure} depicts this situation. We assume that 
$|1>$ corresponds to the state of the detector whereby the analyzer is 
magnetized (spin polarized) in the +z-direction and $|2>$ corresponds to the 
state of the 
detector whereby the analyzer is magnetized in the -z-direction. Thus,
\begin{equation}
|1> \rightarrow \left [ \begin{array}{c}
             1 \\
             0 \\
             \end{array}   \right]
\end{equation}
\begin{equation}
|2> \rightarrow \left [ \begin{array}{c}
             0 \\
             1 \\
             \end{array}   \right]
\end{equation}
Clearly $|1>$ and $|2>$ are orthogonal and the state $|1> + |2>$ 
corresponds to the state
\begin{equation}
|1> + |2> \rightarrow \left [ \begin{array}{c}
             1 \\
             1 \\
             \end{array}   \right] ~,
\end{equation}
which corresponds to spin-polarization in the +x-direction.

Thus, we must find the analyzer polarized in the +x-direction 
after the quanton passes through it.
In other words, the magnetization of the analyzer must be 
sensitive to the passage of a quanton and respond to it.
At present, this is not possible; but the recent discovery of 
control of magnetization via an electric current in InMnAs
\cite{ohno2} may one day lead to a practical paradigm for 
acheiving this.

\section{Conclusions}

In this review article, I have provided a bird's eye view of the 
current status of quantum information science from the perspective of a device 
engineer. This is one of the fastest growing fields judging by the 
number of new journals that are appearing with amazing frequency to 
disseminate many of the important new findings. This field will 
continue to grow and my only regret is that when this article 
appears in print it would probably have been already outdated.

\section{Acknowldegement}

My interest in quantum computing was kindled at the Quantum Devices and
Circuits conference held in Alexandria, Egypt in 1996 which I co-organized.
That conference included presentations by some of the pioneers in this
field including Charles Bennet and Umesh Vazirani. I learnt a lot
from my friend and colleague Prof. Vwani Roychowdhury of UCLA. The quantum dot 
realization of the Toffoli-Fredkin gate is the 
work of my ex-student  Prof. Alexander Balandin of the University of 
California-Riverside during his tenure as a post-doctoral advisee in my group. I 
also acknowledge fruitful discussions with my colleagues Professors
Frazer Williams, Natale Ianno and David Sellmyer of my erstwhile institution, 
the University of Nebraska-Lincoln with which I will always have a special 
connection. I have benefitted from discussions with Prof. David Janes of 
Purdue University and Dr. Berry Jonker of the Naval Research Laboratory
regarding fabrication and MBE growth. The work in spintronics is carried out in 
collaboration with Prof. Marc Cahay of the University of Cincinnati.

My work in quantum computing has been supported by the Nebraska Research 
Initiative in
Quantum Information Processing and the National Science Foundation.

This article is dedicated to the memory of Dr. Rolf Landauer whose work in this
field has taught me much. 

\pagebreak

\begin{center}
{\large \bf Appendix: Rashba Effect in a Quantum Dot}
\end{center}

\begin{figure}[ht]
\epsfxsize=4.2in
\centerline{\epsffile{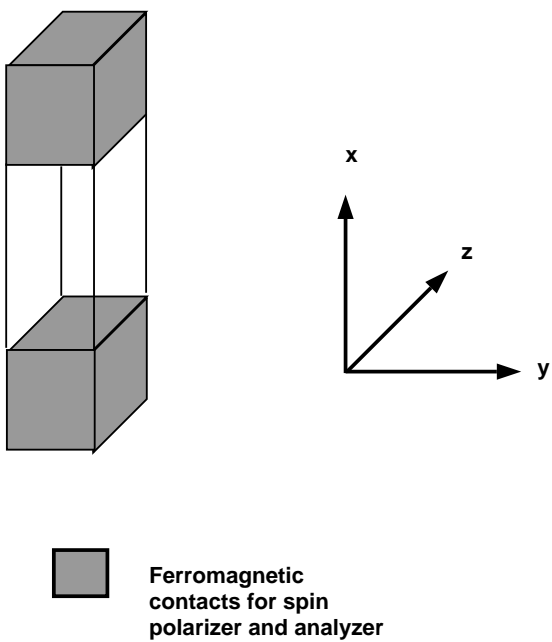}}
\caption{\tt The geometry of a semiconductor quantum dot capped by ferromagnetic 
contacts. The 
contacts induce a magnetic field in the x-direction, which in turn, induces
a Zeeman splitting of the subband levels in the semiconductor quantum dot. The
dot does not have hardwall boundaries. Consequently, the wave functions of the 
spin split levels leak out into the adjoining barrier. The eigenstate of
the spin aligned against the magnetic field has a higher energy and hence 
its wave function leaks out more into the barrier compared to the wave function 
of the spin aligned along the magnetic field. \label{appendix} }
\end{figure}

In this appendix, we will derive the total spin-splitting in a quantum dot 
capped by ferromagnetic contacts.
Consider the system shown in Fig. \ref{appendix}. The ferromagnetic contacts 
give 
rise to an in-built magnetic field in the x-direction which can be quite
strong in realistic structures ($\sim$ 1 Tesla). There may be an 
electric field in the x-direction as well to maintain single electron
occupancy and an electric field in the y-direction to induce the
Rashba effect.

The total Hamiltonian for an electron in the semiconductor layer is
\begin{eqnarray}
H  & = & {{({\vec p} - e{\vec A})^2}\over{2m^*}} -e{\cal E}_x x - e {\cal E}_y y
+ (g/2) \mu_B B_x \sigma_x + H_R \nonumber \\
& = & H_0 + H_R
\end{eqnarray}
where $g$ is the Land\'e g-factor, $e$ is the electronic charge,
$\mu_B$ is the Bohr magneton,
$\sigma_x$ is the x-component of the Pauli spin matrix, and $H_R$ is the  Rashba 
interaction  
given by
\begin{equation}
H_R =  i {{\hbar^2}\over{2m^{*2} c^2}} \nabla V \cdot \left [ {\vec \sigma} 
\times {\vec \nabla} \right ] = {{ e \hbar}\over{2m^{*2} c^2}} {\vec {\cal E}} 
\cdot  {\vec \sigma} \times {\vec p}
\end{equation}
where ${\vec \sigma}$ is the Pauli spin matrix, ${\vec p}$ is the 
momentum operator, and ${\vec {\cal E}}$ is the electric field inducing the 
Rashba effect.

We will neglect the effect of the x-directed electric field  on
the Rashba effect and include only the effect of the y-directed field
which is much larger since the potential applied 
along the x-direction must be smaller than $e/2C$ to maintain Coulomb
blockade. In any case, the x-directed field has no Rashba effect
on x-polarized spins, and we are interested in the energy splitting 
between the +x and -x-polarized spins.
Therefore,
\begin{equation}
H_R = {{ e \hbar}\over{2m^{*2} c^2}} {\cal E}_y \left [\sigma_z p_x - \sigma_x 
p_z \right ]
\end{equation}

The Zeeman term $(g/2) \mu_B B_x \sigma_x$ introduces an Zeeman 
splitting between the +x-polarized spin ($|\uparrow>$) and the 
-x-polarized spin ($|\downarrow>$). If the potential confining
the electron in the semiconductor quantum dot is finite (the 
conduction band offset between the semiconductor and the surrounding
material is finite), then the spatial parts of the ``upspin'' (+x-polarized)
and ``downspin'' (-x-polarized) states are slightly different. The 
wave function of the higher energy state will be spread out a little 
bit more. This is shown in Fig. 1(d). Thus, if the downspin state is at a higher 
energy, then the spatial 
parts of the 
wave functions of the two spin states in the lowest spin-split subband of the 
quantum dot can be written as
\begin{eqnarray}
\phi \uparrow = \left ({{2\sqrt{2}}\over{\sqrt{W_x W_y W_z}}} \right )   sin 
\left ( {{\pi x}\over{W_x}} \right )
  sin \left ( {{\pi y}\over{W_y}} \right ) sin \left ( {{\pi z}\over{W_z}} 
\right ) \nonumber \\
\phi \downarrow = \left ({{2\sqrt{2}}\over{\sqrt{W'_x W'_y W'_z}}} \right )   
sin \left ( {{\pi x}\over{W'_x}} \right )
  sin \left ( {{\pi y}\over{W'_y}} \right ) sin \left ( {{\pi z}\over{W'_z}} 
\right )
\end{eqnarray}
where $W'_x$ $>$ $W_x$, $W'_y$ $>$ $W_y$ and $W'_z$ $>$ $W_z$. These 
widths are larger than the physical dimensions of the quantum dot
since the wave functions will leak out into the barrier as long as the
barrier is not of infinite height. The point of this exercise is to 
show that the spatial parts of the two spin states are different because 
of the Zeeman splitting. This is a critical requirement for the Rashba 
splitting.

Next, we will evaluate the total spin splitting ($\Delta$) which is a
combination of the Zeeman and Rashba splitting. The latter can be modulated by 
the transverse gate potential. 

The time-independent Schr\"odinger equation describing the ground state of
the system
is 
\begin{equation}
\left (H_0 + H_R \right ) \psi = E \psi
\label{expansion}
\end{equation}

We will expand $\psi$ in the basis functions of the two lowest spin-resolved
subband states. We can neglect the higher subband states as long as the 
Rashba spin splitting $\Delta_R$ is much smaller than the energy 
separation between the lowest two subbands in the quantum dot.
If the effective mass is equal to the free electron mass and the 
dimensions of the quantum dot in all directions is about 10 nm,
then the energy separation between the two lowest subbands is 33 meV.
This is obviously much larger than any reasonable Rashba splitting which 
scarcely exceeds 1 meV. Hence, neglecting the higher subbands is
justified.

Hence
\begin{eqnarray}
\psi & = &  a_{\uparrow} \phi \uparrow + a_{\downarrow} \phi \downarrow
\end{eqnarray}

Using the above in Equation \ref{expansion}, we get

\begin{equation}
 \left [ \begin{array}{cc}
             <H_1> + <H_R>_{11} & <H_R>_{12} \\
              <H_R>_{21} & <H_2> + <H_R>_{22} \\
             
\end{array}   \right] 
 \left( \begin{array}{c}
             a_{\uparrow} \\
             a_{\downarrow} \\
             
\end{array}   \right)
=
E_{ground}  \left( \begin{array}{c}
             a_{\uparrow} \\
             a_{\downarrow} \\
             \end{array}   \right) ~,
\end{equation}
where $<H_1>$ = $<\phi \uparrow |H_0|\phi \uparrow>$, $<H_2>$ = $<\phi 
\downarrow |H_0|\phi \downarrow>$, $<H_R>_{11}$ = $<\phi \uparrow |H_R|\phi 
\uparrow>$, $<H_R>_{22}$ = $<\phi \downarrow |H_R|\phi \downarrow>$,
$<H_R>_{12}$ = $<\phi \uparrow |H_R|\phi \downarrow>$, and $<H_R>_{21}$ = $<\phi 
\downarrow |H_R|\phi \uparrow>$. 

Diagonalizing the above Hamiltonian, we get
that the total splitting between the upspin and downspin states is
\begin{eqnarray}
E_{\downarrow} - E_{\uparrow} & = & 2 \sqrt{\left ( {{<H_1> - <H_2> + <H_R>_{11} 
- <H_R>_{22}}\over{2}} \right )^2 + <H_R>_{12}<H_R>_{21}  } \nonumber \\
& = & 2 \sqrt{ \left ( {{g \mu_B B}\over{2}} + {{ e \hbar}\over{2m^{*2} c^2}} 
{\cal E}_y <p_x> \right )^2 + \left |{{e \hbar}\over{2m^{*2} c^2}} {\cal E}_y  
<p_z> \right |^2 }
\end{eqnarray}
where $<p_x>$ = $<\phi \uparrow | -i \hbar (\partial/\partial x | \phi \uparrow 
>$ = $<\phi \downarrow | -i \hbar (\partial/\partial x | \phi \downarrow >$ = 0
and $<p_z>$ = $<\phi \uparrow | -i \hbar (\partial/\partial z | \phi \downarrow 
>$ = ${{8 i \hbar}\over{\sqrt{W_z W'_z}}}cos^2 \left ({{\pi}\over{2}}  {{ 
W_x}\over{W'_x}} \right)$.

Therefore the total splitting is
\begin{equation}
\Delta = E_{\downarrow} - E_{\uparrow} = 2 \sqrt{ \left ( {{g \mu_B B}\over{2}}  
 \right )^2 +  {\cal E}_y^2  {{16 e^2\hbar^4}\over{ m^{*4}c^4 W_x W'_x}} cos^4 
\left ({{\pi W_x}\over{2 W'_x}} \right ) }
\end{equation}

The last term under the radical is the Rashba effect which can be varied
by the electric field ${\cal E}_y$. Note that this term would have 
vanished if $W_x$ = $W'_x$, that is, if the spatial parts of the upspin
and downspin wave functions were identical. Here, we have made the 
spatial parts different by using a finite potential barrier and a magnetic 
field to raise the energy of the downspin state above that of the upspin
state.

We can estimate the magnitude of the Rashba splitting in 
realistic systems. Assuming $W_x$ = 10 nm, $W'_x$ = 11 nm,
${\cal E}_y$ = 10$^{7}$ V/m, $B$ = 1 Tesla, $g$ = 2, we find 
that increasing the electric field from 10$^{7}$ V/m to twice 
that value increases the Rashba splitting by about 1.5 $\mu$eV.
Thus, the modulation of the splitting is very small.

\pagebreak


\begin{thebibliography}{10}
\bibitem{vwani2}
The author was made aware of this proof by Professor Vwani Roychowdhury.
\end{thebibliography}

\end{document}